\documentclass[aps,prl,twocolumn,superscriptaddress, amsmath,amssymb]{revtex4-2}

\usepackage{diagbox}
\usepackage{ragged2e}
\usepackage{array}
\usepackage{booktabs}
\usepackage{graphicx}
\usepackage{dcolumn}
\usepackage{bm}
\usepackage[colorlinks, linkcolor=black, anchorcolor=black, citecolor=black]{hyperref}

\providecommand{\enquote}[1]{#1}

\begin{document}

\preprint{APS/123-QED}

\title{Temperature-Induced Superconductivity Enhancement under Large Exchange Field}

\author{Xusheng Wang}
\email{wang-xs23@mails.tsinghua.edu.cn}
\affiliation{%
State Key Laboratory of Low-Dimensional Quantum Physics, Department of Physics, Tsinghua University, Beijing 100084, China}
\author{Lianyi He}
\affiliation{%
State Key Laboratory of Low-Dimensional Quantum Physics, Department of Physics, Tsinghua University, Beijing 100084, China}
\author{Shuai-hua Ji}
\affiliation{%
State Key Laboratory of Low-Dimensional Quantum Physics, Department of Physics, Tsinghua University, Beijing 100084, China}
\affiliation{%
Frontier Science Center for Quantum Information, Beijing 100084, China}

\date{\today}

\begin{abstract}

Through a comprehensive free energy analysis, we demonstrate that finite temperature can simultaneously weaken superconductivity and mitigate spin polarization induced depairing, leading to potential non-monotonic temperature-dependent behaviors in superconductors subjected to large exchange fields. Remarkably, superconductivity can be counterintuitively enhanced by temperature when the Zeeman energy exceeds the superconducting order parameter, owing to the competition between thermal and magnetic effects. We propose that multiband effect offers one possible microscopic route for this temperature-induced enhancement and demonstrate it explicitly within a two-band superconducting model. A detailed parameter analysis identifies the conditions under which this phenomenon emerges, suggesting that temperature-enhanced superconductivity may be observable in materials such as $\mathrm{MgB_2}$ and $\mathrm{FeSe}$ through transport and tunneling measurements.

\end{abstract}

\maketitle

The behavior of superconductors under large exchange fields has been a central topic in condensed matter physics for decades. The zero-temperature critical exchange field $H_c$ was first determined independently by Clogston and Chandrasekhar \cite{clogston1962upper,chandrasekhar1962note}, now known as the Pauli limit. The finite temperature extension of this theory was later developed by Sarma through a free energy analysis \cite{sarma1963influence,maki1966effect,fulde1973high}, which revealed that the transition from the homogeneous superconducting state to the normal state is first-order at low temperatures and second-order at higher temperatures. Subsequent studies demonstrated that an inhomogeneous superconducting state with finite-momentum pairing, the Fulde-Ferrell-Larkin-Ovchinnikov (FFLO) state, can become energetically favorable slightly beyond the Pauli limit \cite{fulde1964superconductivity,larkin1965nonuniform}. However, realization of the FFLO state requires exceptionally clean systems to sustain the spatial modulation of the order parameter. These theoretical predictions have been supported by experimental observations in superconducting thin films subjected to in-plane magnetic fields \cite{tedrow1970experimental,meservey1970magnetic,meservey1975tunneling,tedrow1977supercooling,tedrow1977measurement,kasahara2020evidence,cho2021evidence,wan2023orbital}.


We focus on the temperature-induced non-monotonic behavior of superconductivity under strong exchange fields. In conventional understanding, increasing temperature suppresses superconductivity, causing the order parameter to decrease monotonically. However, superconductors containing magnetic impurities or exhibiting spontaneous ferromagnetism challenge this expectation. Sarma and de Gennes theoretically demonstrated that, in superconductors with an appropriate concentration of magnetic impurities, the temperature-dependent polarization of the impurities can lead to non-monotonic behavior: the $H_c$ can be enhanced by thermal effects at low temperatures regimes \cite{de1966some,bennemann1966anomalous}. This prediction was supported by experiments on superconducting dilute rare-earth alloys \cite{crow1966transition,crow1967superconductivity,bennemann1969nature}. Furthermore, several independent studies have reported temperature-induced non-monotonic or reentrant superconductivity in various ferromagnetic superconductors \cite{ishikawa1977destruction,fertig1977destruction,woolf1979superconducting,eisaki1994competition,crespo2009evolution,paramanik2013reentrant,tran2024reentrant}, which were explained in our recent work by incorporating temperature-dependent ferromagnetism \cite{wang2025unified}.

In this Letter, we investigate both one-band and two-band superconducting models applicable to arbitrary order parameter symmetries and finite temperatures. By analyzing the free energy, we show that temperature can simultaneously weaken superconductivity and reduce the depairing effect of exchange fields in any clean superconductor. The competition between these two effects can give rise to non-monotonic temperature dependence of the superconducting order parameter when the Zeeman energy exceeds the superconducting order parameter, resulting in a stronger suppression of spin polarization. Furthermore, we explicitly examine temperature-induced superconductivity enhancement within a two-band framework and identify the parameter regimes in which this phenomenon emerges.

For a two-band superconducting film subjected to an external in-plane magnetic field $H$, which acts purely as an exchange field, the Hamiltonian is expressed as

\begin{align}
\hat{H} &= \sum_{i=1,2}\sum_{\mathbf{k},\sigma} \left(\xi_i(\mathbf{k}) + \sigma H \right) 
\hat{c}^\dagger_{i\sigma}(\mathbf{k}) \hat{c}_{i\sigma}(\mathbf{k}) + \notag \\
& \quad \sum_{i,j=1,2}\sum_{\mathbf{k},\mathbf{k}'} U_{ij}(\mathbf{k},\mathbf{k}')
\hat{c}^\dagger_{i\uparrow}(\mathbf{k}) 
\hat{c}^\dagger_{i\downarrow}(-\mathbf{k})
\hat{c}_{j\downarrow}(-\mathbf{k}') 
\hat{c}_{j\uparrow}(\mathbf{k}'),
\end{align}

\noindent where $\hat{c}^\dagger_{i\sigma}$ ($\hat{c}_{i\sigma}$) represents the $\sigma$ spin electron creation (annihilation) operator in $i^{\mathrm{th}}$ band, and $U_{ij}$ denotes the pairing interaction strength between $i^{\mathrm{th}}$ and $j^{\mathrm{th}}$ bands. Within the mean-field and weak-coupling approximations, the order parameter amplitudes satisfy the relation $\Delta_1/\Delta_{2}=\Delta_{10}/\Delta_{20}$, which simplifies the computation. Under these conditions, the free energy $F(H,T)$ can be derived in closed form (see Supplementary Material (SM) for details).

{\small
\begin{align}\label{free_energy}
& F = \sum_{i}\int_0^{2\pi} \mathrm{d}\phi \left[\chi^2(\phi) N_i \Delta_i^2 \left( \ln \frac{\Delta_i}{\Delta_{i0}} - \frac{1}{2} \right) - \frac{2N_i}{\beta} \int_0^{\infty} \mathrm{d}\xi \, \right. \notag \\
& \left. \ln \Biggl( 1 + 2 \cosh(\beta H) e^{-\beta \sqrt{\xi^2 + \Delta_i^2\chi^2(\phi)}}
+ e^{-2\beta \sqrt{\xi^2 + \Delta_i^2\chi^2(\phi)}} \Biggr)\right],
\end{align}
}

\noindent where $N_i$ is the density of states near fermi level for the $i^{\mathrm{th}}$ band, $\Delta_i$ the corresponding superconducting order parameter with $\Delta_{i0}=\Delta_i(H=0,T=0)$, and $\beta = 1/k_B T$. The function $\chi(\phi)$ characterizes the order parameter anisotropy ($\Delta_i(\phi) = \Delta_i\chi(\phi)$), with $\chi(\phi)=1$ for isotropic s-wave and $\chi(\phi)=\cos 2\phi$ for anisotropic d-wave pairing. For simplicity, only the s-wave case is discussed in the main text, while results for d-wave and other forms of $\chi(\phi)$ are provided in the SM. Minimizing the free energy $F(H,T)$ yields the self-consistent $\Delta(H,T)$, enabling quantitative analysis of the temperature-induced superconductivity enhancement.

For computational convenience, Eq.~(\ref{free_energy}) is expressed in dimensionless form as $f(x)=F/(N_1\Delta_{10}^2)$ with $x=\Delta_i/\Delta_{i0}$. The other dimensionless parameters are defined as

\begin{equation}\label{dimensionless_parameters}
    \alpha=\frac{\Delta_{20}}{\Delta_{10}},\quad \gamma=\frac{N_1}{N_2}, \quad h=\frac{H}{\Delta_{10}}, \quad t=\frac{k_BT}{\Delta_{10}}.
\end{equation}

The explicit form of the dimensionless two-band free energy is given in the SM. Because $\Delta_1$ and $\Delta_2$ are interchangeable, we take $\Delta_1$ to be the smaller order parameter in the following discussion without loss of generality, implying $\alpha \geq 1$. The two-band case naturally reduces to the one-band limit when $\alpha = 1$, which will be discussed first as an illustrative example. To analyze the free energy difference between the normal state ($x=0$) and the superconducting state ($x \neq 0$), we define

\begin{equation}\label{difference_energy}
    f_{ns}(x,h,t) = f(x,h,t)-f(0,h,t).
\end{equation}

\begin{figure}[ht]
\centering
\includegraphics[width=1.0\linewidth]{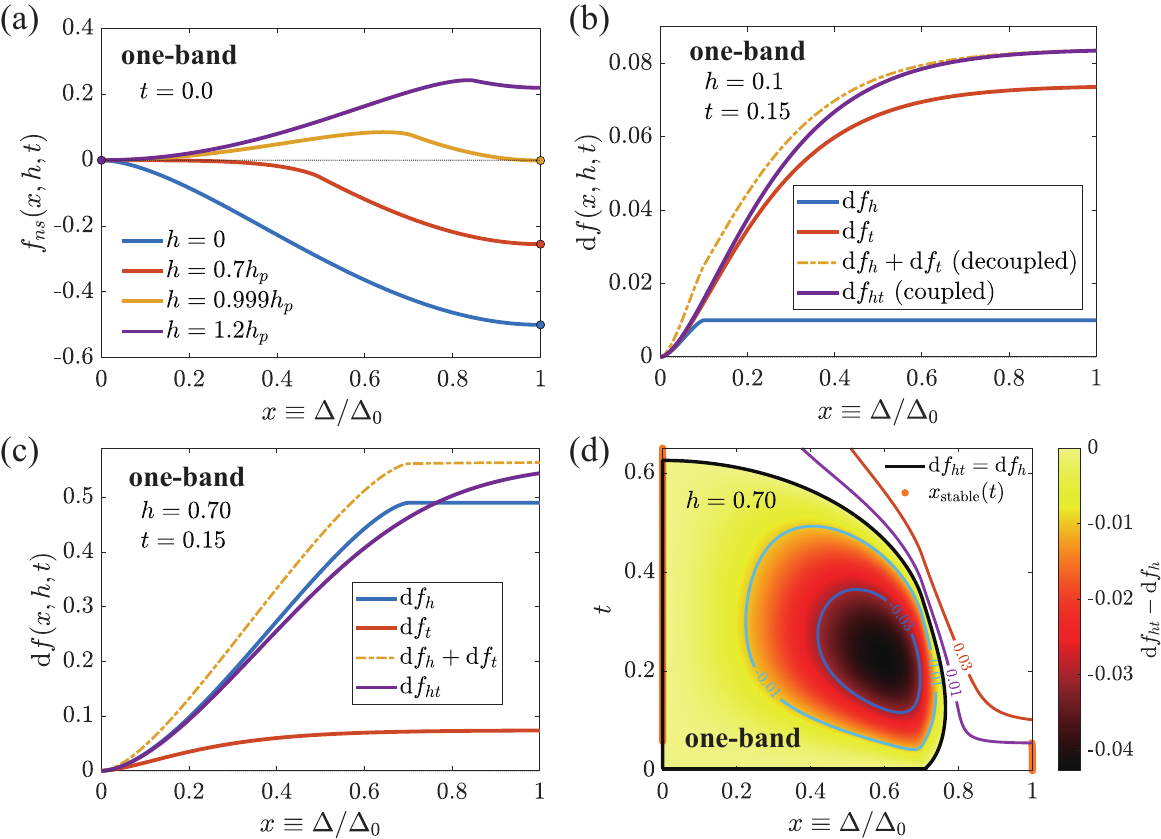}
\caption{
Free energy analysis for one-band s-wave superconductors.
(a) Landscapes of $f_{ns}(x)$ at $t=0$ for $h=0.0$, $0.7h_p$, $0.999h_p$, and $1.2h_p$.
(b,c) Corresponding profiles of $\mathrm{d}f_{h}(x)$, $\mathrm{d}f_{t}(x)$, $\mathrm{d}f_{h}(x)+\mathrm{d}f_{t}(x)$, and $\mathrm{d}f_{ht}(x)$ for (b) $h=0.10$, $t=0.15$ and (c) $h=0.70$, $t=0.15$.
(d) Temperature evolution map of $\mathrm{d}f_{ht}(x)-\mathrm{d}f_{h}(x)$ at $h=0.70$.
The solid black line marks the condition $\mathrm{d}f_{ht}(x)=\mathrm{d}f_{h}(x)$, orange points $x_{\mathrm{stable}}$ denote the equilibrium order parameter, and the color scale represents the amplitude of $\mathrm{d}f_{ht}(x)-\mathrm{d}f_{h}(x)$, corresponding to the negative of the energy saved by temperature effects.
Contour lines of $\mathrm{d}f_{ht}(x)-\mathrm{d}f_{h}(x)$ illustrate the distribution of the energy saving.
Regions where $\mathrm{d}f_{ht}(x)-\mathrm{d}f_{h}(x)\ge 0$ are shown as transparent but remain traceable through their positive contour levels.}
\label{free_energy_difference}
\end{figure}

Since $f(0,h,t)$ is independent of the order parameter, minimizing $f_{ns}(x,h,t)$ yields the same self-consistent solution for $x$. The difference is that $f_{ns}(x,h,t)$ defines the free energy of the normal state $f_{ns}(0,h,t)$ as a constant zero reference, independent of $h$ and $t$. Fig.~\ref{free_energy_difference}(a) shows the evolution of $f_{ns}$ at zero temperature for various $h$ in a one-band s-wave superconductor, where the minima of the curves correspond to the equilibrium order parameters. The order parameter exhibits an abrupt drop from $x=1$ to $x=0$ once $h$ exceeds the Pauli limit $h_p = 1/\sqrt{2}$, consistent with previous studies \cite{clogston1962upper,chandrasekhar1962note,sarma1963influence}. Notably, $f_{ns}$ increases with $h$ and displays pronounced $x$ dependence. To quantify the variation of $f_{ns}$ induced by the exchange field and temperature, we offset $f_{ns}(x,h,t)$ by its zero-field and zero-temperature reference value $f_{ns}(x,h=0,t=0)$, as illustrated by the blue curve in Fig.~\ref{free_energy_difference}(a), leading to the definition

\begin{equation}\label{offset_difference_energy}
    \mathrm{d}f(x,h,t) = f_{ns}(x,h,t)-f_{ns}(x,h=0,t=0).
\end{equation}

Based on Eq.~(\ref{offset_difference_energy}) and the discussion above, $\mathrm{d}f(x,h,t)\equiv \mathrm{d}f_{ht}$, $\mathrm{d}f(x,h,t{=}0)\equiv \mathrm{d}f_{h}$, and $\mathrm{d}f(x,h{=}0,t)\equiv \mathrm{d}f_{t}$ represent the lifted free energies associated with the combined effects of finite $h$ and $t$, the pure spin polarization and the pure thermal effects, respectively. These quantities, calculated for finite $h$ and $t$ and shown in Fig.~\ref{free_energy_difference}(b) and \ref{free_energy_difference}(c), illustrate the competition between the temperature-induced reduction of spin polarization and the thermal suppression of superconductivity.

For a small $h=0.1$ with relatively large $t=0.15$, Fig.~\ref{free_energy_difference}(b) shows that $\mathrm{d}f_{h}+\mathrm{d}f_{t}\geq\mathrm{d}f_{ht}\geq\mathrm{d}f_{h}$ holds for all $x$, indicating that temperature reduces the free energy via coupling with the exchange field but cannot offset the increase arising from superconductivity suppression. In contrast, for $h=0.7$ near $h_p$ and $t=0.15$, $\mathrm{d}f_{h}\geq\mathrm{d}f_{ht}$ appears at small and intermediate $x$, as shown in Fig.~\ref{free_energy_difference}(c). To examine this temperature-induced energy saving in detail, the evolution of $\mathrm{d}f_{h}-\mathrm{d}f_{ht}$ with temperature is mapped in Fig.~\ref{free_energy_difference}(d). The solid black line in Fig.~\ref{free_energy_difference}(d) is consistent with the results in Fig.~\ref{free_energy_difference}(c), showing that free energy reduction occurs only when the Zeeman energy generally exceeds the order parameter. Therefore, at sufficiently large exchange fields, a finite temperature can diminish spin polarization more effectively than it suppresses superconductivity, resulting in a net energy gain relative to the zero-temperature case (see SM for details). As temperature rises, this saved energy initially increases and then rapidly decreases after reaching a maximum at a moderate temperature.

For a one-band s-wave superconductor, the order parameter remains constant ($x=1$) as $h$ increases and abruptly drops to $x=0$ once $h\geq h_p$. Consequently, thermal effects cannot reduce the free energy, and no temperature-induced enhancement of superconductivity appears. This motivates the search for scenarios in which the order parameter can be partially suppressed before superconductivity vanishes, allowing the system to accommodate a large Zeeman energy relative to its order parameter. One such possibility is a highly anisotropic order parameter, briefly discussed in the SM, while another arises in two-band superconductors \cite{barzykin2007gapless,barzykin2009magnetic,he2009stable}.

\begin{figure}[ht]
\centering
\includegraphics[width=1.0\linewidth]{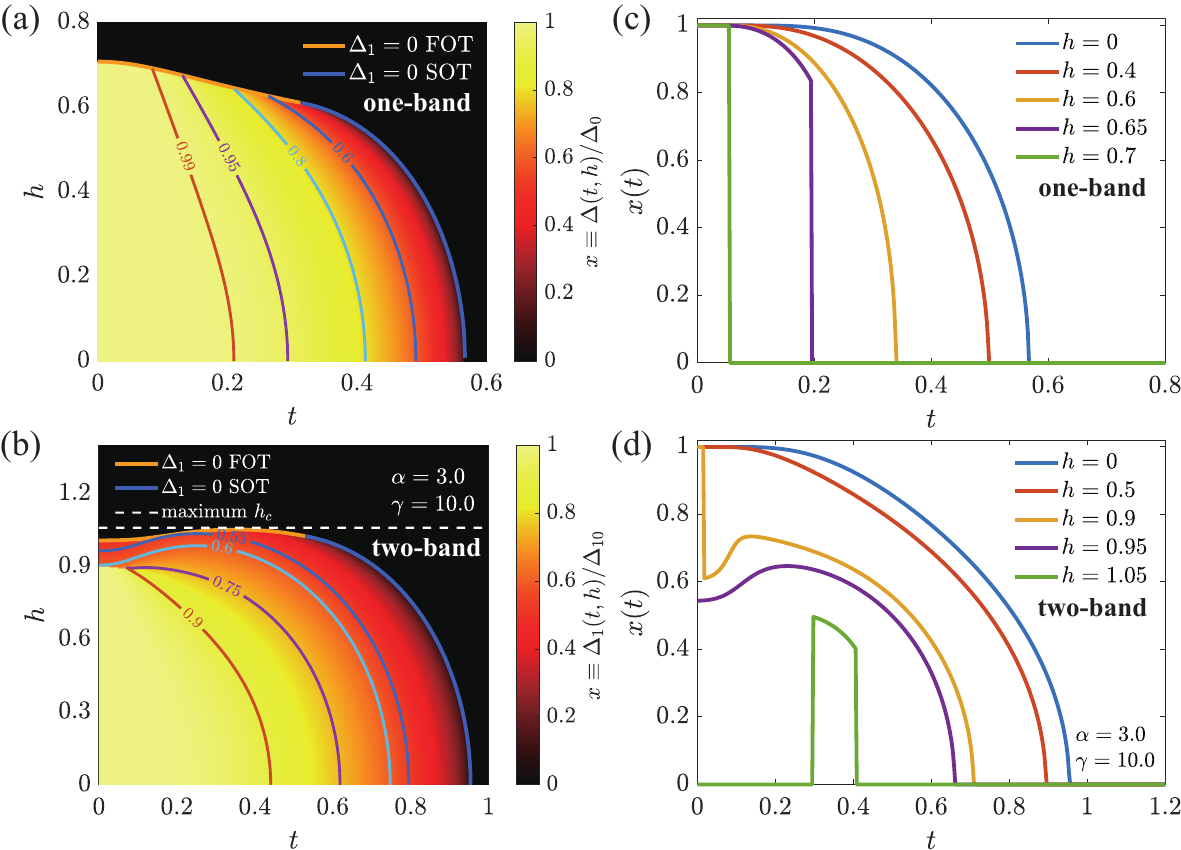}
\caption{
Superconductivity under large exchange fields for one-band and two-band s-wave systems.
(a,b) $h$-$t$ phase diagrams for (a) one-band and (b) two-band case with specific parameters.
The first-order and second-order transition boundaries (FOT and SOT) are denoted by orange and blue solid lines, respectively.
Color scales indicate the magnitude of the order parameter $x$, and contour lines highlight the monotonic and non-monotonic superconducting features.
The same color scales and line conventions are used in the following figures.
(c,d) Temperature dependence of the order parameter $x(t)$ at various exchange fields for (c) one-band and (d) two-band systems.
}
\label{h_t_phase_diagram_one_two}
\end{figure}

The superconducting phase diagrams of one-band and two-band s-wave systems are presented for comparison. For the one-band case, the $h$-$t$ phase diagram in Fig.~\ref{h_t_phase_diagram_one_two}(a) shows that the critical field $h_c$ reaches its maximum at zero temperature and gradually decreases with increasing temperature. The contour lines of the order parameter extend monotonically from the upper-left to the lower-right corner, indicating that superconductivity is not enhanced by thermal effects. At a fixed in-plane magnetic field, the order parameter $x$ decreases steadily with temperature [Fig.~\ref{h_t_phase_diagram_one_two}(c)], further confirming the absence of temperature-induced superconductivity enhancement.

In contrast, temperature-induced superconductivity enhancement appears in the two-band system under specific parameter conditions. As highlighted by the dashed white line in Fig.~\ref{h_t_phase_diagram_one_two}(b), the maximum of $h_c(t)$ occurs at a finite temperature rather than at zero, confirming the presence of enhanced superconductivity. At zero temperature, the order parameter is partially suppressed to $x\approx 0.5$ near the critical field where $h_c>x$, allowing thermal effects to more effectively mitigate the depairing caused by the exchange field. Consequently, the contour lines of the order parameter in the $h$-$t$ phase diagram exhibit distinct non-monotonic features. Moreover, Fig.~\ref{h_t_phase_diagram_one_two}(d) shows that the temperature dependence of $x$ varies with the applied field $h$. For relatively small $h$, the order parameter remains $x=1$ at zero temperature and decreases monotonically with increasing temperature, consistent with the one-band case. When $h$ becomes large enough to induce a FOT satisfying $h>x$, temperature-related non-monotonic behavior emerges.

For a relatively large $h$ that triggers a FOT at finite temperature rather than at zero, the order parameter exhibits a sequence of behaviors: an abrupt drop, a gradual recovery, and then suppression with further temperature increase as illustrated by the $h=0.9$ case in Fig.~\ref{h_t_phase_diagram_one_two}(d). With increasing $h$, when the FOT already occurs at zero temperature but remains below $h_c$, the order parameter first increases and then decreases with temperature, as seen for $h=0.95$ in Fig.~\ref{h_t_phase_diagram_one_two}(d). Most interestingly, when $h$ slightly exceeds $h_c$ at zero temperature, a reentrant superconducting behavior emerges, where the order parameter becomes non-zero only within an intermediate temperature range. In practice, additional non-monotonic temperature dependences of the order parameter can appear for certain parameter sets. The corresponding results are provided in the SM.

\begin{figure}[ht]
\centering
\includegraphics[width=1.0\linewidth]{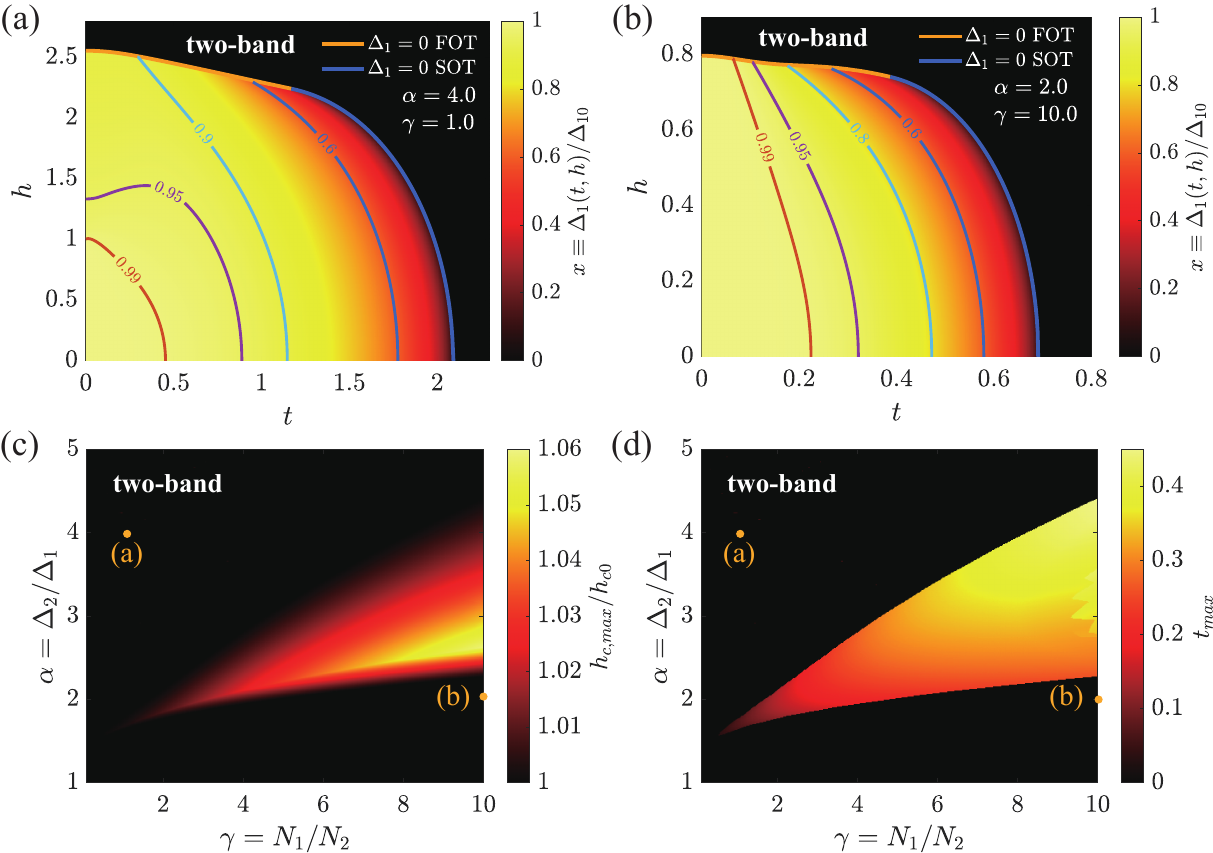}
\caption{
$\alpha$-$\gamma$ parameter dependence of superconductivity enhancement.
(a,b) $h$-$t$ phase diagrams for two-band systems with (a) large $\alpha$ and small $\gamma$, and (b) small $\alpha$ and large $\gamma$.
(c,d) Maps of (c) $h_{c,max}/h_{c0}$ and (d) $t_{max}$ in the $\alpha$-$\gamma$ parameter space.
The color scales in (c) and (d) represent the amplitudes of $h_{c,max}/h_{c0}$ and $t_{max}$, respectively.
The parameter points corresponding to (a) and (b) are indicated by orange markers.
}
\label{parameter_dependence_hc_enhancement}
\end{figure}

The properties of two-band systems depend strongly on the ratio of order parameters $\alpha$ and the ratio of densities of states $\gamma$, as defined in Eq.~(\ref{dimensionless_parameters}). For two-band systems with large $\alpha$ and small $\gamma$, superconductivity is dominated by the larger order parameter $\Delta_2$, resembling a one-band system characterized by $\Delta_2$. Conversely, for small $\alpha$ ($\alpha \approx 1$) and large $\gamma$, the behavior is governed by the smaller order parameter $\Delta_1$, also effectively mimicking a one-band system. As shown in Fig.~\ref{parameter_dependence_hc_enhancement}(a,b), the $h$-$t$ phase diagrams of these two limiting cases closely resemble that of the one-band system in Fig.~\ref{h_t_phase_diagram_one_two}(a). However, the $\Delta_2$-dominated two-band system can exhibit weak non-monotonic temperature dependence near a SOT at $h \approx x$, as indicated by the $x=0.95$ contour line in Fig.~\ref{h_t_phase_diagram_one_two}(b), while the $\Delta_1$-dominated system does not, since it always satisfies $h<x$. Moreover, both types of one order parameter dominated systems show no temperature-induced enhancement near the critical field, as evidenced by the maximum $h_c$ occurring at zero temperature, confirming that superconductivity enhancement is highly sensitive to the parameters $\alpha$ and $\gamma$.

The enhancement of superconductivity is characterized by two key features: the temperature-induced non-monotonic behavior of the order parameter and $h_c$ enhancement. Because the non-monotonic evolution of the order parameter is difficult to quantify directly, we focus on the parameter dependence of the $h_c$ enhancement. We define the critical exchange field at zero temperature as $h_{c0}$ and the maximum critical field as $h_{c,max}$, which occurs at temperature $t_{max}$. Accordingly, temperature-induced superconductivity enhancement is identified when $h_{c,max}/h_{c0} > 1$. The maps of $h_{c,max}/h_{c0}$ and $t_{c,max}$ in the $\alpha$-$\gamma$ parameter space were calculated to reveal this dependence. As shown in Fig.~\ref{parameter_dependence_hc_enhancement}(c), the enhancement of $h_c$ appears only within an appropriate region of the parameter space, where $\alpha$ and $\gamma$ are roughly comparable. In other words, the parameters should avoid approaching the one-band limits, thereby retaining the intrinsic two-band characteristics. The corresponding $t_{max}$ map in Fig.~\ref{parameter_dependence_hc_enhancement}(d) further indicates that the maximal superconductivity enhancement always occurs at a relatively small but non-zero temperature.

As discussed in Fig.~\ref{free_energy_difference}, energy saving occurs only when the Zeeman energy exceeds the superconducting order parameter, which is always accompanied by a partially suppressed order parameter. For a two-band superconductor at zero temperature, the critical field $h_{c0}$ induces a FOT, at which the homogeneous order parameter abruptly drops to zero from a finite value. To evaluate the order parameter immediately before the transition, denoted as $x(h_{c0}^-)$, we introduce a small field increment $\mathrm{d}h = 1\times10^{-5}$ and define

\begin{equation}
    x(h_{c0}^-)=x(h_{c0}-\mathrm{d}h).
\end{equation}

\begin{figure}[ht]
\centering
\includegraphics[width=1.0\linewidth]{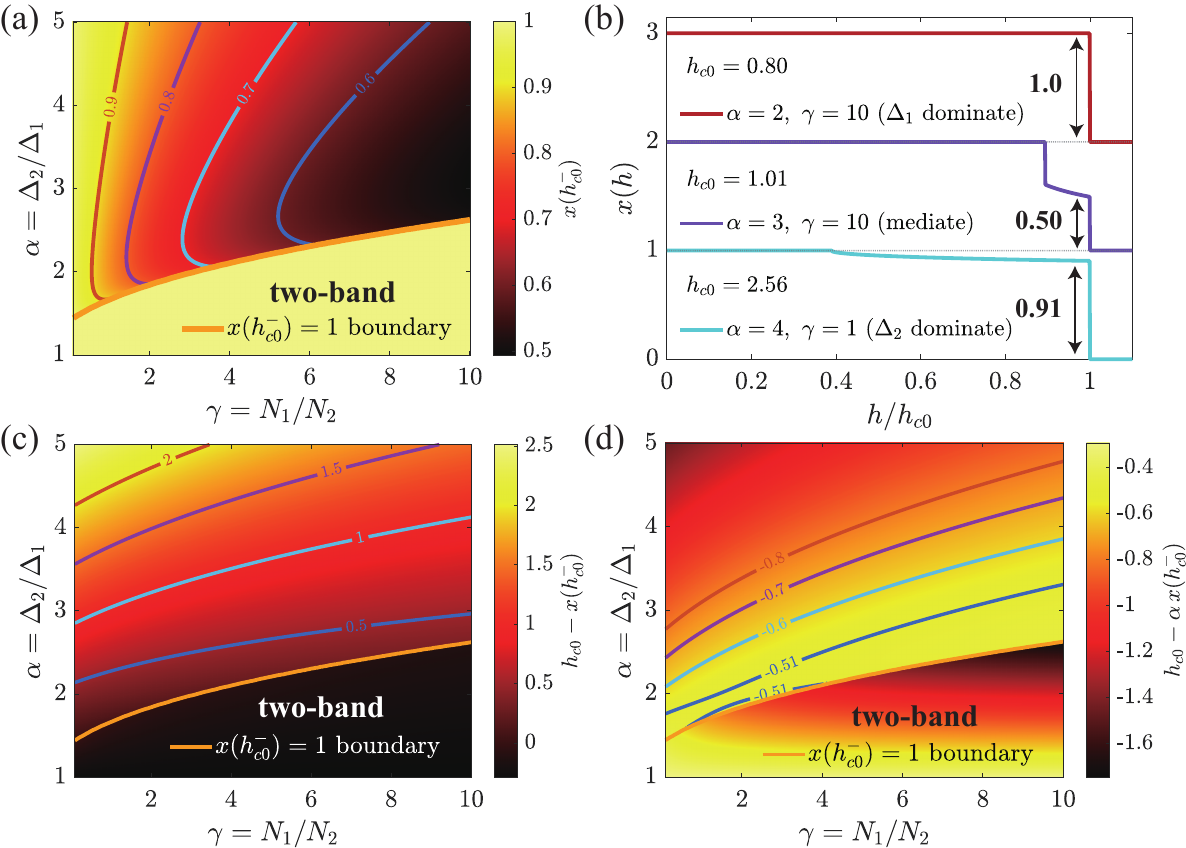}
\caption{
$\alpha$-$\gamma$ parameter dependence of the relation between $x(h_{c0}^-)$ and $h_{c0}$ at zero temperature.
(a) Map of $x(h_{c0}^-)$.
(b) Order parameter profiles $x(h)$ for different $\alpha$ and $\gamma$ values at zero temperature. Each curve is vertically offset by $1$ for clarity, and the drop near $h_{c0}$ is marked by double-sided arrows.
(c,d) Maps of (c) $h_{c0}-x(h_{c0}^-)$ and (d) $h_{c0}-\alpha x(h_{c0}^-)$.
The color scales in (a), (c), and (d) represent the amplitudes of $x(h_{c0}^-)$, $h_{c0}-x(h_{c0}^-)$, and $h_{c0}-\alpha x(h_{c0}^-)$, respectively.
Contour lines of $h_{c0}-\alpha x(h_{c0}^-)$ below the boundary defined by $x(h_{c0}^-)=1$ are omitted for clarity.
}
\label{parameter_dependence_order_parameter_drop}
\end{figure}

The $\alpha$ and $\gamma$ parameter dependence of $x(h_{c0}^-)$ is shown in Fig.~\ref{parameter_dependence_order_parameter_drop}(a). The contour lines display non-monotonic variation primarily along the $\alpha$ direction, indicating that the minimum of $x(h_{c0}^-)$ occurs at large $\gamma$ and intermediate $\alpha$, located near the middle-right region of the map. This behavior can be further understood from the order parameter profiles in Fig.~\ref{parameter_dependence_order_parameter_drop}(b). For small $\alpha$ and large $\gamma$ systems, where the superconductivity is dominated by $\Delta_1$, only a single FOT appears, similar to the one-band case. For intermediate $\alpha$ with large $\gamma$, where the stronger order parameter partially protects the weaker one, an additional FOT emerges at relatively low $h$, leading to significant suppression of the order parameters. For large $\alpha$ with relatively small $\gamma$, where the stronger order parameter effectively protects the weaker one, only a SOT occurs at low $h$, resulting in minimal suppression. Additional details are provided in the SM.

The contour line distributions of $x(h_{c0}^-)$ are broadly consistent with the $h_c$ enhancement map in Fig.~\ref{parameter_dependence_hc_enhancement}(c), indicating that superconductivity enhancement is always associated with partial suppression of the order parameters. Minor discrepancies arise from the dominance of $\Delta_2$ at high fields in systems with large $\alpha$. To further compare the Zeeman energy and order parameters at the critical field, we calculated the maps of $h_{c0}-x(h_{c0}^-)$ and $h_{c0}-\alpha x(h_{c0}^-)$, as shown in Fig.~\ref{parameter_dependence_order_parameter_drop}(c,d). In Fig.~\ref{parameter_dependence_order_parameter_drop}(c), increasing $\alpha$ and decreasing $\gamma$ make $h_{c0}$ significantly larger than the weaker order parameter $x=\Delta_1/\Delta_{10}$, enabling non-monotonic order parameter behavior near $h=x$, as seen in Fig.~\ref{parameter_dependence_hc_enhancement}(a). By contrast, Fig.~\ref{parameter_dependence_order_parameter_drop}(d) shows that the Zeeman energy $h_{c0}$ remains smaller than the stronger component of order parameter $\alpha x$, and the system behaves effectively as a one-band superconductor dominated by $\Delta_2$ when $\alpha$ is too large and $\gamma$ too small, suppressing any $h_c$ enhancement.

In summary, we have studied a two-band model applicable to arbitrary temperatures and order-parameter symmetries. Through a systematic free energy analysis of the one-band case, we showed that finite temperature simultaneously suppresses superconductivity and spin polarization induced depairing in any clean superconductor and can enhance superconductivity when the Zeeman energy exceeds the superconducting order parameter. Possible microscopic origins include spin-orbit-parity coupling and two-band effects, where previous works \cite{xie2020spin,barzykin2009magnetic} have reported slight temperature-induced $h_c$ enhancement in $h$-$t$ phase diagrams, but did not explicitly identify or emphasize this phenomenon. In the present study, we analyzed two-band s-wave superconductors in detail and demonstrated that temperature-induced non-monotonic order parameter behavior and $h_c$ enhancement serve as signatures of enhanced superconductivity. We found that this enhancement is highly sensitive to the parameters $\alpha$ and $\gamma$: only systems with large $\gamma$ and intermediate $\alpha$ exhibit clear $h_c$ enhancement, whereas non-monotonic order-parameter behavior can arise whenever $\alpha$ and $\gamma$ are not small.

Our theoretical predictions, the temperature-induced $h_c$ enhancement and the non-monotonic temperature dependence of the order parameter, can be directly tested by transport and tunneling measurements on multiband superconducting thin films under large in-plane magnetic field. Two-band s-wave systems such as $\mathrm{MgB_2}$ \cite{zehetmayer2013review,liu2001beyond,szabo2001evidence,iavarone2002two} and $\mathrm{NbSe_2}$ \cite{yokoya2001fermi,kiss2007charge,noat2015quasiparticle,dvir2018spectroscopy}, as well as anisotropic (possibly d-wave) two-band superconductors including $\mathrm{CeCoIn_5}$ \cite{petrovic2001heavy,rourke2005spectroscopic,seyfarth2008multigap,allan2013imaging} and $\mathrm{FeSe}$ \cite{khasanov2010evolution,bourgeois2016thermal,sun2017gap,sprau2017discovery}, provide ideal platforms for such verification. In $\mathrm{MgB_2}$, tunneling measurements \cite{iavarone2002two} reveal spatial variations in the ratio of densities of states $\gamma$, offering a natural setting to test our s-wave predictions. Moreover, previous studies have shown that both the densities of states ratio $\gamma$ and the order parameter ratio $\alpha$ can be tuned by substrate-induced hole doping \cite{lin2023real,huang2024experimental}, making $\mathrm{FeSe}$ a particularly promising candidate for experimentally validating our theoretical framework. 

\textit{Acknowledgments.--} We acknowledge X.-Y. Zhou, Z.-C. Zhang, L.-C Ji, Y.-Q. Yan, J.-B. Liu, and M. Shu for stimulating discussions. This work is financially supported by
Quantum Science and Technology-National Science and Technology Major Project (Grants No. 2023ZD0300500) and the National Natural Science Foundation of China (Grants No. 11427903, 52388201 and 12375136).\\


\nocite{*}

\providecommand{\noopsort}[1]{}\providecommand{\singleletter}[1]{#1}%

\pagebreak
\appendix
\onecolumngrid
\setcounter{figure}{0}
\renewcommand*{\thefigure}{S\arabic{figure}}
\setcounter{equation}{0}
\renewcommand*{\theequation}{S\arabic{equation}}

\begin{widetext}
\section{\Large{Supplementary Material}}
\section{I. Free energy derivation and computational details for the two-band model}

Considering an arbitrary pairing interaction, the Hamiltonian in the main text can be rewritten as~\cite{he2009stable}
\begin{align}
\hat{H} &= 
\sum_{i=1,2}\sum_{\mathbf{k},\sigma} 
\xi_{i\sigma}(\mathbf{k}) 
\hat{c}^\dagger_{i\sigma}(\mathbf{k}) 
\hat{c}_{i\sigma}(\mathbf{k})
+ 
\sum_{i,j=1,2}\sum_{\mathbf{k},\mathbf{k}'} 
U_{ij}(\mathbf{k},\mathbf{k}')
\hat{c}^\dagger_{i\uparrow}(\mathbf{k})
\hat{c}^\dagger_{i\downarrow}(-\mathbf{k})
\hat{c}_{j\downarrow}(-\mathbf{k}')
\hat{c}_{j\uparrow}(\mathbf{k}'),
\quad 
\xi_{i\sigma}(\mathbf{k}) = \xi_i(\mathbf{k}) + \sigma H,
\label{hamiltonian}
\end{align}

\noindent where all symbols are defined in the main text. The pairing interaction $U_{ij}(\mathbf{k},\mathbf{k}')$ generally takes the form

\begin{equation}
U_{ij}(\mathbf{k},\mathbf{k}') = -g_{ij}\,\chi(\phi)\chi'(\phi'),
\end{equation}

\noindent
where $g_{12} = g_{21}$ follows from the symmetry of the interaction, and $\chi(\phi)$ represents the appropriate irreducible representation of the pairing symmetry~\cite{barzykin2007gapless}. According to the order parameter symmetry, the s-wave and d-wave cases correspond to $\chi(\phi) = 1$ and $\chi(\phi) = \cos 2\phi$, respectively. The superconducting order parameter is then defined as

\begin{align}
\Delta_i(\mathbf{k}) &= \sum_{j=1,2}\sum_{\mathbf{k}'} 
U_{ij}\,
\langle \hat{c}_{j\downarrow}(-\mathbf{k}') \hat{c}_{j\uparrow}(\mathbf{k}') \rangle
= \Delta_i \chi(\phi), \notag \\
\Delta_i &= -\sum_{j=1,2}\sum_{\mathbf{k}'} 
g_{ij}\chi(\mathbf{k}')\, 
\langle \hat{c}_{j\downarrow}(-\mathbf{k}') \hat{c}_{j\uparrow}(\mathbf{k}') \rangle.
\label{eq:gap}
\end{align}

In practice, $\Delta_i$ can be conveniently chosen real and positive without loss of generality. Under the mean-field approximation, the Hamiltonian in Eq.~(\ref{hamiltonian}) simplifies to

\begin{align}
\hat{H} &=
\frac{g_{22}\Delta_1^2 + g_{11}\Delta_2^2 - 2g_{12}\Delta_1\Delta_2}
     {g_{11}g_{22} - g_{12}^2}
+ \sum_{i=1,2}\sum_{\mathbf{k}}
\left[
\begin{pmatrix}
\hat{c}^\dagger_{i\uparrow}(\mathbf{k}) & \hat{c}^\dagger_{i\downarrow}(-\mathbf{k})
\end{pmatrix}
\begin{pmatrix}
\xi_{i\uparrow}(\mathbf{k}) & \Delta_i(\mathbf{k}) \\
\Delta_i(\mathbf{k}) & -\xi_{i\downarrow}(-\mathbf{k})
\end{pmatrix}
\begin{pmatrix}
\hat{c}_{i\uparrow}(\mathbf{k}) \\
\hat{c}_{i\downarrow}(-\mathbf{k})
\end{pmatrix}
+ \xi_{i\downarrow}(-\mathbf{k})
\right].
\label{hamiltonian2}
\end{align}

Using the standard Bogoliubov transformation, following a procedure similar to that described in Ref.~\cite{wang2025unified}, the Hamiltonian in Eq.~(\ref{hamiltonian2}) can be diagonalized into the form

\begin{equation}
\label{bogoliubov}
\hat{H} =
\frac{g_{22}\Delta_1^2 + g_{11}\Delta_2^2 - 2g_{12}\Delta_1\Delta_2}
     {g_{11}g_{22} - g_{12}^2}
+ \sum_{i=1,2}\sum_{\mathbf{k}} \left[ \xi_i(\mathbf{k}) - E_i(\mathbf{k}) \right]
+ \sum_{i=1,2}\sum_{\mathbf{k}}\sum_{s=\pm}
E_{is}(\mathbf{k})\,
\hat{\alpha}^\dagger_{is}(\mathbf{k}) \hat{\alpha}_{is}(\mathbf{k}),
\end{equation}

\noindent where $E_{is}(\mathbf{k}) = E_i(\mathbf{k}) + s H$, and $E_i(\mathbf{k}) = \sqrt{\Delta_i^2(\mathbf{k}) + \xi_i^2(\mathbf{k})}$. Here, $\hat{\alpha}^\dagger_{is}$ ($\hat{\alpha}_{is}$) denotes the creation (annihilation) operator for Bogoliubov quasiparticles in the $i^{\mathrm{th}}$ band, which still obey fermionic statistics. The free energy $F$ is obtained from the partition function, $F = -\tfrac{1}{\beta}\ln \mathcal{Z}$, and can be written as

\begin{equation}
F = \frac{g_{22}\Delta_1^2+g_{11}\Delta_2^2-2g_{12}\Delta_1\Delta_2}{g_{11}g_{22}-g_{12}^2}
+ \sum_{i=1,2}\sum_{\mathbf{k}} \left( \xi_i({\mathbf{k}}) - E_i({\mathbf{k}}) \right)
- \frac{1}{\beta} \sum_{i=1,2} \sum_{\mathbf{k}} \sum_{s=\pm} 
\ln \left( 1 + e^{-\beta E_{is}(\mathbf{k})} \right),
\end{equation}

Assuming that both bands possess constant densities of states $N_i$ within the pairing interaction window, the free energy can be expressed in integral form as
\begin{equation}
\label{free_energy_supple}
F =
\frac{g_{22}\Delta_1^2 + g_{11}\Delta_2^2 - 2g_{12}\Delta_1\Delta_2}
     {g_{11}g_{22} - g_{12}^2}
+ \sum_{i=1,2} 2N_i
\int \frac{\mathrm{d}\hat{\mathbf{k}}}{4\pi}
\int_0^{E_c} \mathrm{d}\xi_i
\left[
\xi_i(\mathbf{k}) - E_i(\mathbf{k})
- \frac{1}{\beta}
\sum_{s=\pm}
\ln\!\left(1 + e^{-\beta E_{is}(\mathbf{k})}\right)
\right],
\end{equation}

\noindent where $E_c$ is the cutoff energy. In the limit $E_c \gg \Delta_i(\mathbf{k})$, the temperature-independent terms can be integrated analytically, and Eq.~(\ref{free_energy_supple}) simplifies to

\begin{align}
F &=
\frac{g_{22}\Delta_1^2 + g_{11}\Delta_2^2 - 2g_{12}\Delta_1\Delta_2}
     {g_{11}g_{22} - g_{12}^2}
- \sum_{i=1,2} N_i
\int \frac{\mathrm{d}\hat{\mathbf{k}}}{4\pi}
\frac{\Delta_i^2 \chi^2(\mathbf{k})}{2}
\left[1 + \ln\!\frac{4E_c^2}{\Delta_i^2 \chi^2(\mathbf{k})}\right] \notag \\
&\quad
- \sum_{i=1,2} \frac{2N_i}{\beta}
\int \frac{\mathrm{d}\hat{\mathbf{k}}}{4\pi}
\int_0^\infty \mathrm{d}\xi_i
\ln\!\left(
1 + 2\cosh(\beta H)e^{-\beta E_i(\mathbf{k})}
+ e^{-2\beta E_i(\mathbf{k})}
\right).
\label{free_energy_for_gap_eq}
\end{align}

In Eq.~(\ref{free_energy_for_gap_eq}), it should be noted that the upper limit of the temperature-dependent integral has been extended to infinity, which is justified by the convergence of the integral and the sufficiently large cutoff energy $E_c$. The self-consistent gap equations, which the order parameters must satisfy, are obtained by applying the stationary conditions $\partial F / \partial \Delta_1 = \partial F / \partial \Delta_2 = 0$. Focusing on the zero-temperature and zero-field case, we obtain

\begin{align}
\Delta_{10} N_1 \int \frac{\mathrm{d}\hat{\mathbf{k}}}{4\pi}
\frac{\chi^2(\mathbf{k})}{2}
\ln\!\frac{4E_c^2}{\Delta_{10}^2 \chi^2(\mathbf{k})}
&= \frac{g_{22}\Delta_{10} - g_{12}\Delta_{20}}
       {g_{11}g_{22} - g_{12}^2}, \notag \\
\Delta_{20} N_2 \int \frac{\mathrm{d}\hat{\mathbf{k}}}{4\pi}
\frac{\chi^2(\mathbf{k})}{2}
\ln\!\frac{4E_c^2}{\Delta_{20}^2 \chi^2(\mathbf{k})}
&= \frac{g_{11}\Delta_{20} - g_{12}\Delta_{10}}
       {g_{11}g_{22} - g_{12}^2}.
\label{gap_equations}
\end{align}

By eliminating the cutoff energy $E_c$ using Eq.~(\ref{gap_equations}), the free energy can be rewritten in a cutoff-independent form as

\begin{align}
F &=
\frac{g_{12}\Delta_{10}\Delta_{20}}
     {g_{11}g_{22} - g_{12}^2}
\left(
\frac{\Delta_1}{\Delta_{10}} - \frac{\Delta_2}{\Delta_{20}}
\right)^{\!2}
+ \sum_{i=1,2} N_i
\int \frac{\mathrm{d}\hat{\mathbf{k}}}{4\pi}
\chi^2(\mathbf{k}) \Delta_i^2
\left(
\ln\!\frac{\Delta_i}{\Delta_{i0}} - \tfrac{1}{2}
\right) \notag \\
&\quad
- \frac{2}{\beta}
\sum_{i=1,2} N_i
\int \frac{\mathrm{d}\hat{\mathbf{k}}}{4\pi}
\int_0^\infty \mathrm{d}\xi_i
\ln\!\left[
1 + 2\cosh(\beta H)e^{-\beta \sqrt{\Delta_i^2(\mathbf{k}) + \xi_i^2}}
+ e^{-2\beta \sqrt{\Delta_i^2(\mathbf{k}) + \xi_i^2}}
\right].
\label{free_energy_calcu_ori}
\end{align}

To facilitate numerical calculations, the free energy in Eq.~(\ref{free_energy_calcu_ori}) is further simplified by introducing the following dimensionless quantities:
\begin{equation}
f(x_1,x_2) \equiv \frac{F}{N_1\Delta_{10}^2}, \quad
\lambda_{ij} = g_{ij} N_j, \quad
x_1 = \frac{\Delta_1}{\Delta_{10}}, \quad
x_2 = \frac{\Delta_2}{\Delta_{20}}, \quad
t = \frac{k_B T}{\Delta_{10}}, \quad
h = \frac{H}{\Delta_{10}}, \quad
\alpha = \frac{\Delta_{20}}{\Delta_{10}}, \quad
\gamma = \frac{N_1}{N_2}.
\label{dimless2}
\end{equation}

Since all energy-related quantities are normalized by $\Delta_{10}$, we do not explicitly distinguish $h$ and $T$ from their dimensionless forms $\delta$ and $t$ in the following discussions. Substituting the above definitions into Eq.~(\ref{free_energy_calcu_ori}), the reduced free energy becomes

\begin{align}
f(x_1,x_2)
&= \frac{\lambda_{12}\alpha}{\lambda_{11}\lambda_{22}-\lambda_{12}\lambda_{21}}
(x_1 - x_2)^2
+ \int \frac{\mathrm{d}\hat{\mathbf{k}}}{4\pi} \chi^2(\mathbf{k})
\left[
x_1^2\!\left(\ln x_1 - \tfrac{1}{2}\right)
+ \frac{\alpha^2 x_2^2}{\gamma}\!\left(\ln x_2 - \tfrac{1}{2}\right)
\right] \notag \\
&\quad
- 2t \int \frac{\mathrm{d}\hat{\mathbf{k}}}{4\pi}
\int_0^\infty \mathrm{d}\epsilon\,
\ln\!\left[
1 + 2\cosh(h/t)
  e^{-\sqrt{\epsilon^2 + x_1^2\chi^2(\mathbf{k})}/t}
+ e^{-2\sqrt{\epsilon^2 + x_1^2\chi^2(\mathbf{k})}/t}
\right] \notag \\
&\quad
- \frac{2t}{\gamma}
\int \frac{\mathrm{d}\hat{\mathbf{k}}}{4\pi}
\int_0^\infty \mathrm{d}\epsilon\,
\ln\!\left[
1 + 2\cosh(h/t)
  e^{-\sqrt{\epsilon^2 + \alpha^2 x_2^2 \chi^2(\mathbf{k})}/t}
+ e^{-2\sqrt{\epsilon^2 + \alpha^2 x_2^2 \chi^2(\mathbf{k})}/t}
\right].
\label{fx1x2}
\end{align}

Once $h$ and $t$ are specified, $f(x_1,x_2)$ depends solely on the reduced order parameters $x_1, x_2 \in [0,1]$. In this parameter space, the global minimum, local minima, and maxima of $f(x_1,x_2)$, which automatically satisfy Eq.~(\ref{gap_equations}), represent the stable, metastable, and unstable states, respectively. Since this work focuses on equilibrium superconducting properties, only the global minimum solutions are discussed. A detailed analysis of metastable and unstable branches, including their thermodynamic roles, will be reported elsewhere.

Identifying the extrema, particularly the global minimum of the free energy surface $f(x_1,x_2)$ in the two-dimensional parameter space, while simultaneously accounting for the momentum direction and thermal convolutions, can be computationally demanding. To reduce this numerical complexity, we employ the weak-coupling approximation, a standard assumption in BCS theory, where $\lambda_{ij}=g_{ij}N_j\ll1$ \cite{bardeen1957theory}. This approximation, well justified for conventional superconductors, significantly simplifies the analysis of Eq.~(\ref{fx1x2}). In this limit, the prefactor of the first term in the free energy expression Eq.~(\ref{fx1x2}) should be

\begin{equation}
\frac{\lambda_{12}\alpha}{\lambda_{11}\lambda_{22}-\lambda_{12}\lambda_{21}}
\sim \alpha\,\mathcal{O}(\lambda_{ij}^{-1}) \gg 1.
\end{equation}

This condition implies that $x_1$ and $x_2$ should be approximately equal in order to minimize the free energy $f(x_1,x_2)$, allowing us to reduce the problem to a single variable by setting $x = x_1 = x_2$. Furthermore, analyzing the coupled gap equations in the weak-coupling limit yields

\begin{equation}
\lambda_{22} - \lambda_{11}
+ \lambda_{21}\frac{\Delta_{1}}{\Delta_{2}}
- \lambda_{12}\frac{\Delta_{2}}{\Delta_{1}} = 0.
\label{Delta_1_Delta_2}
\end{equation}

As shown in Eq.~(\ref{Delta_1_Delta_2}), the ratio $\Delta_2/\Delta_1$ is independent of both temperature and external magnetic field, and can be determined by solving the corresponding quadratic equation as

\begin{equation}
\alpha = \frac{\Delta_{20}}{\Delta_{10}} = \frac{\Delta_{2}}{\Delta_{1}}
= \frac{\lambda_{22} - \lambda_{11} + \sqrt{(\lambda_{11} - \lambda_{22})^2 + 4\lambda_{12}\lambda_{21}}}{2\lambda_{12}}.
\label{constant_ratio}
\end{equation}

The constant ratio $\Delta_2 / \Delta_1$ is consistent with previous theoretical results~\cite{barzykin2007gapless}.  
Accordingly, the free energy in Eq.~(\ref{fx1x2}) can be reduced to a single-variable form:

\begin{align}
f(x)
&= \int \frac{\mathrm{d}\hat{\mathbf{k}}}{4\pi} \chi^2(\mathbf{k})
\left[
x^2\!\left(\ln x - \tfrac{1}{2}\right)
+ \frac{\alpha^2 x^2}{\gamma}\!\left(\ln x - \tfrac{1}{2}\right)
\right] \notag \\
&\quad
- 2t \int \frac{\mathrm{d}\hat{\mathbf{k}}}{4\pi}
\int_0^\infty \mathrm{d}\epsilon\,
\ln\!\left[
1 + 2\cosh(h/t)
  e^{-\sqrt{\epsilon^2 + x^2\chi^2(\mathbf{k})}/t}
+ e^{-2\sqrt{\epsilon^2 + x^2\chi^2(\mathbf{k})}/t}
\right] \notag \\
&\quad
- \frac{2t}{\gamma}
\int \frac{\mathrm{d}\hat{\mathbf{k}}}{4\pi}
\int_0^\infty \mathrm{d}\epsilon\,
\ln\!\left[
1 + 2\cosh(h/t)
  e^{-\sqrt{\epsilon^2 + \alpha^2 x^2 \chi^2(\mathbf{k})}/t}
+ e^{-2\sqrt{\epsilon^2 + \alpha^2 x^2 \chi^2(\mathbf{k})}/t}
\right].
\label{fx_approx}
\end{align}

Throughout this work, both in the main text and in the present Supplementary Material, the equilibrium superconducting state is obtained by minimizing the free energy given in Eq.~(\ref{fx_approx})).

\newpage
\section{II. Validation of the weak-coupling limit under finite coupling strengths}

As demonstrated above, the weak-coupling limit leads to a constant ratio $\Delta_2/\Delta_1$ that is independent of both temperature and exchange field. To further validate this approximation, we solved the system using the full free-energy expression [Eq.~(\ref{fx1x2})] for finite coupling strengths and compared the results with those obtained from the weak-coupling form [Eq.~(\ref{fx_approx})]. The procedures used to identify the global minima in both cases are described below.

Since the free energy landscapes $f(x_1,x_2)$ and $f(x)$ in Eqs.~(\ref{fx1x2}) and (\ref{fx_approx}) typically exhibit multiple local minima, special care was taken to ensure accurate identification of the global minimum. For the full two-variable free energy [Eq.~(\ref{fx1x2})], a coarse-grid scan with $200\times200$ points over the domain $x_1, x_2 \in [0,1]$ was first performed to locate the approximate minimum region. Subsequently, the lowest-energy grid point, together with several randomly selected seed points (to guarantee robustness), was used as the initial guesses for the \texttt{fmincon} optimization function in \textsc{MATLAB} to refine the minima with high precision. The lowest-energy solution was then retained for further analysis. Finally, the resulting minimum was compared with $f(x_1,x_2)$ evaluated at the boundaries ($x_1, x_2 = 0, 1$) to eliminate potential edge effects. A similar procedure was applied to the single-variable free energy function [Eq.~(\ref{fx_approx})] to ensure consistency and accuracy. The reliability of the entire algorithm was verified by setting $\alpha=1$, which corresponds to the well-known one-band limit, and has been shown in the main text [Fig. 2(a) and Figures in Section IV]. 

As discussed above, the coupling strength is effectively reflected by the prefactor of the first term in Eq.~(\ref{fx1x2}). To quantitatively evaluate the influence of finite coupling, we define a dimensionless parameter representing the coupling strength as

\begin{equation}
\mathrm{PF} =
\frac{\lambda_{12}}
     {\lambda_{11}\lambda_{22} - \lambda_{12}\lambda_{21}}
\sim \frac{1}{\lambda_{ij}} \gg 1.
\end{equation}

The parameter $\mathrm{PF}$ serves as an indicator of the effective coupling strength: larger values of $\mathrm{PF}$ correspond to weaker coupling. By tuning $\mathrm{PF}$, we simulated systems with different coupling strengths. The field-dependent order parameters $\Delta_i(h)$ were first calculated at zero temperature ($t=0$).

\begin{figure}[ht]
  \centering
  \includegraphics[width=1.0\linewidth]{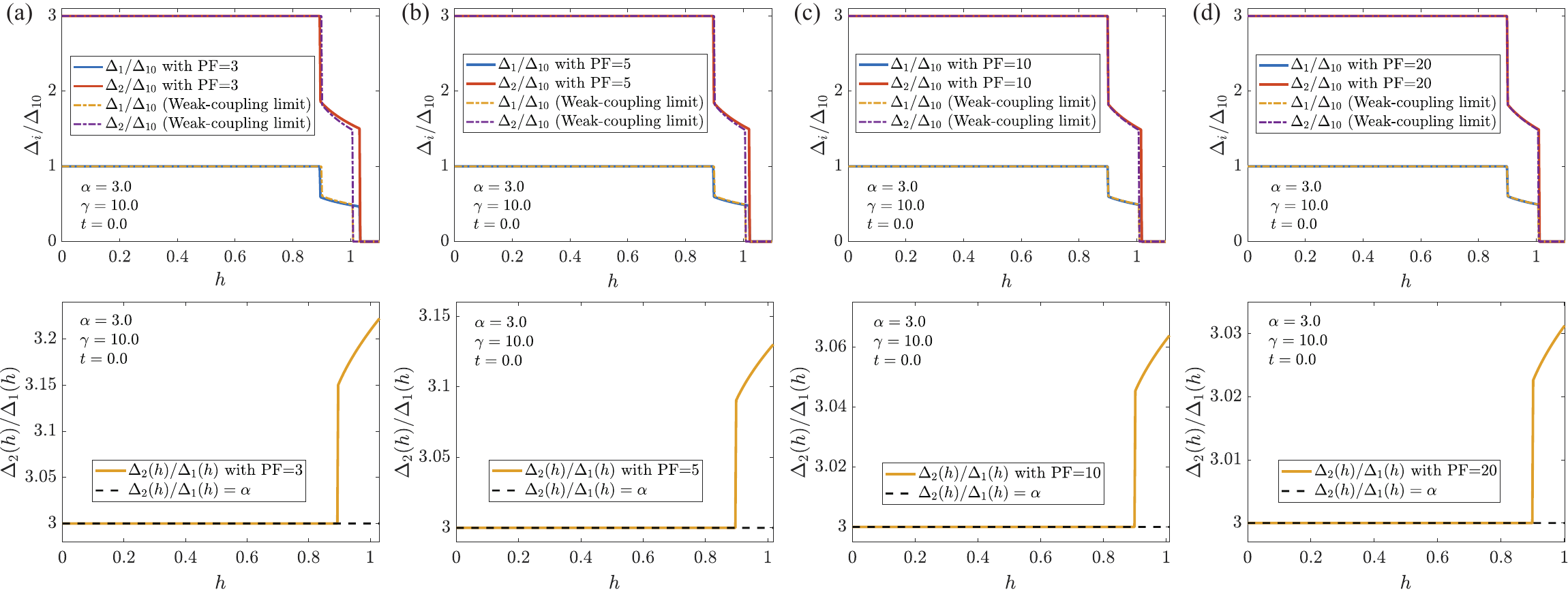}
  \caption{
  Comparison between the weak-coupling approximation and finite-coupling calculations for the field dependence of $\Delta_i(h)$ and $\Delta_2(h)/\Delta_1(h)$. 
  (a-d) Results obtained from the full free energy [Eq.~(\ref{fx1x2})] for different coupling strengths:
  (a) $\mathrm{PF}=3$, (b) $\mathrm{PF}=5$, (c) $\mathrm{PF}=10$, and (d) $\mathrm{PF}=20$.
  The corresponding weak-coupling results from Eq.~(\ref{fx_approx}) are also shown in panels (a-d) for comparison.}
  \label{Comparison_weak_coupling_delta_h}
\end{figure}

As shown in Fig.~\ref{Comparison_weak_coupling_delta_h}(a)-\ref{Comparison_weak_coupling_delta_h}(d), the order parameters obtained within the weak-coupling approximation exhibit only minor deviations from those computed at finite coupling. For stronger couplings ($\mathrm{PF}=3$ and $\mathrm{PF}=5$), small discrepancies around $5\%$ appear mainly in the high-field regime [Figs.~\ref{Comparison_weak_coupling_delta_h}(a,b)]. For weaker couplings ($\mathrm{PF}=10$ and $\mathrm{PF}=20$), the results are nearly indistinguishable from the weak-coupling limit, and the proportionality $\Delta_2(h)/\Delta_1(h)=\alpha$ is preserved with deviations below $2\%$. Thus, the weak-coupling relation $\Delta_2/\Delta_1=\alpha$ remains quantitatively accurate even for moderately strong coupling.

We next evaluated the temperature-dependent order parameters $\Delta_i(t)$ at zero field ($h=0$), as shown in Fig.~\ref{Comparison_weak_coupling_delta_t}. The same conclusion holds: the weak-coupling approximation provides an excellent quantitative description even at relatively large coupling strengths.

\begin{figure}[ht]
  \centering
  \includegraphics[width=1.0\linewidth]{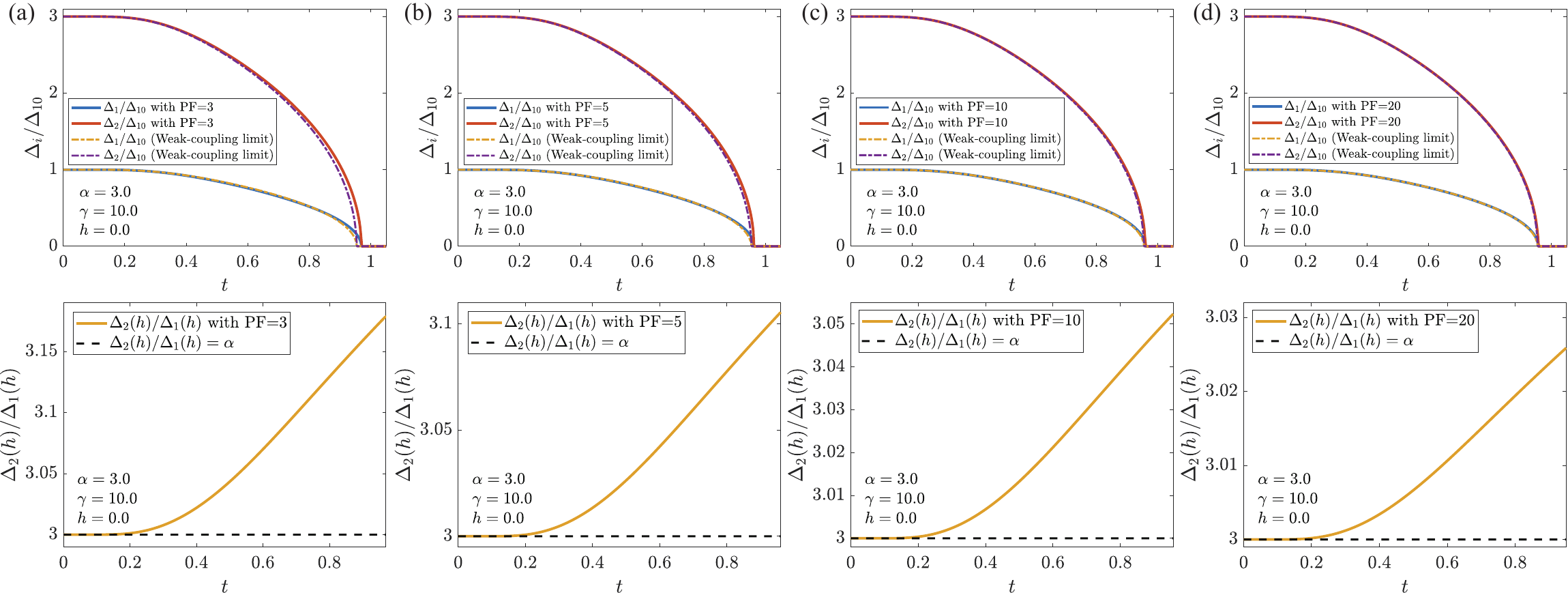}
  \caption{
  Comparison between the weak-coupling approximation and finite-coupling calculations for the temperature dependence of $\Delta_i(t)$ and $\Delta_2(t)/\Delta_1(t)$. 
  (a-d) Results from the full free energy [Eq.~(\ref{fx1x2})] for different coupling strengths:
  (a) $\mathrm{PF}=3$, (b) $\mathrm{PF}=5$, (c) $\mathrm{PF}=10$, and (d) $\mathrm{PF}=20$.
  The corresponding weak-coupling results from Eq.~(\ref{fx_approx}) are also shown in panels (a-d) for comparison.}
  \label{Comparison_weak_coupling_delta_t}
\end{figure}

Because our study focuses on temperature-induced superconductivity enhancement, we also investigated the influence of finite coupling on the critical field $h_c=H_c/\Delta_{10}$. Figure~\ref{Comparison_weak_coupling_hc_t} presents the results for a relatively strong coupling case ($\mathrm{PF}=3$).

\begin{figure}[ht]
  \centering
  \includegraphics[width=0.75\linewidth]{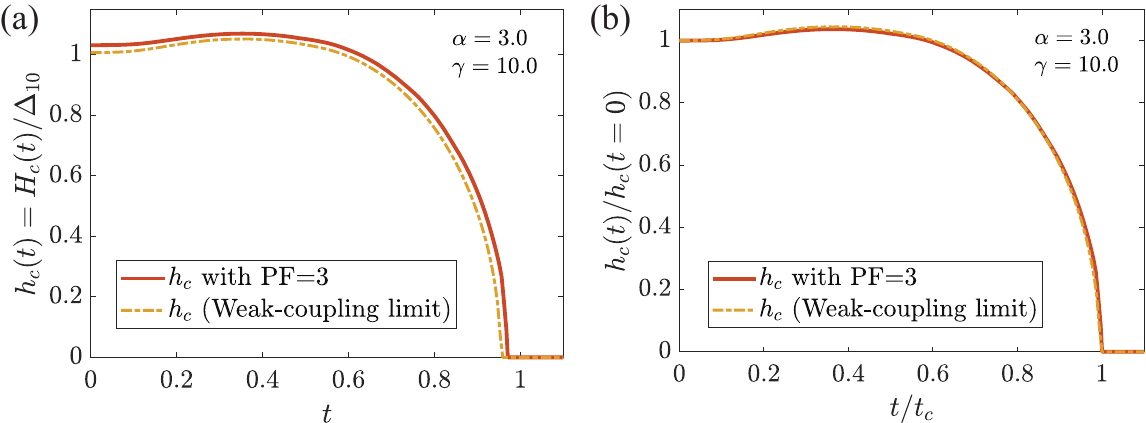}
  \caption{
  Comparison between the weak-coupling approximation and finite-coupling calculations for the critical field $h_c(t)$.
  (a) Calculated $h_c(t)$ curves and 
  (b) normalized $h_c(t)/h_c(0)$ curves obtained from the full free energy [Eq.~(\ref{fx1x2})] with $\mathrm{PF}=3$ and from the weak-coupling approximation [Eq.~(\ref{fx_approx})].}
  \label{Comparison_weak_coupling_hc_t}
\end{figure}

As shown in Fig.~\ref{Comparison_weak_coupling_hc_t}(a), only a slight deviation is observed between the $h_c(t)$ curves obtained from $\mathrm{PF}=3$ and that from the weak-coupling approximation. Remarkably, when both the critical field and the temperature are normalized by their zero-temperature and zero-field values, respectively, that is, when plotting $h_c(t)/h_c(0)$ as a function of $t/t_c$, the two curves almost perfectly overlap [Fig.~\ref{Comparison_weak_coupling_hc_t}(b)]. This result demonstrates that finite coupling acts primarily as a proportional scaling factor on the magnitudes of $h_c(t)$ and $t_c$ without altering the intrinsic temperature dependence. The nearly identical normalized $h_c(t)/h_c(0)$ curves confirm that the $h_c$ enhancement discussed in the main text remains robust over a wide range of coupling strengths.

In summary, the weak-coupling relation $\Delta_2/\Delta_1=\alpha$ is quantitatively exact for sufficiently weak coupling ($\mathrm{PF}\geq 10$) and remains approximately valid even for relatively strong coupling ($\mathrm{PF}=3$). Moreover, the normalized critical field $h_c(t)/h_c(0)$ is nearly insensitive to variations in coupling strength, remaining unchanged even at $\mathrm{PF}=3$. Therefore, the weak-coupling approximation adopted throughout this work is well justified for two-band superconductors, including those with moderate coupling strengths.

\newpage
\section{III. Detailed free energy analysis under exchange field}

As discussed in the main text (see Fig.~1), the thermal effect can simultaneously weaken superconductivity and mitigate the depairing caused by spin polarization. When the exchange field becomes comparable to or exceeds the superconducting condensation energy, thermal excitation reduces spin polarization more effectively than it suppresses superconductivity, leading to a counterintuitive temperature-induced enhancement of superconductivity. In this section, we perform a detailed free energy analysis for both one-band and two-band systems, adopting the s-wave case for computational simplicity.

\subsection{A. One-band free energy analysis}

In the main text [Fig.~1(d)], we presented the temperature evolution map of $\mathrm{d}f_{ht}-\mathrm{d}f_{h}$ for $h=0.70$. Here, we extend this analysis to a wider range of exchange fields. The quantities $\mathrm{d}f_{ht}(x)$, $\mathrm{d}f_{h}(x)$, and $\mathrm{d}f_{t}(x)$ are defined as

\begin{equation}
    \mathrm{d}f_{ht}(x)\equiv f_{ns}(x,h,t),\quad  \mathrm{d}f_{h}(x)\equiv f_{ns}(x,h,t=0),\quad  \mathrm{d}f_{t}(x)\equiv f_{ns}(x,h=0,t),
\end{equation}

\noindent where $f_{ns}(x)=f(x)-f(x=0)$ is the superconducting free energy measured relative to the normal state, which is set to zero. Fig.~\ref{detailed_free_energy_one_band} shows the detailed temperature evolution of $f_{ns}(x,h,t)$ and $\mathrm{d}f_{ht}-\mathrm{d}f_{h}$ for various exchange fields.

\begin{figure}[ht]
  \centering
  \includegraphics[width=0.95\linewidth]{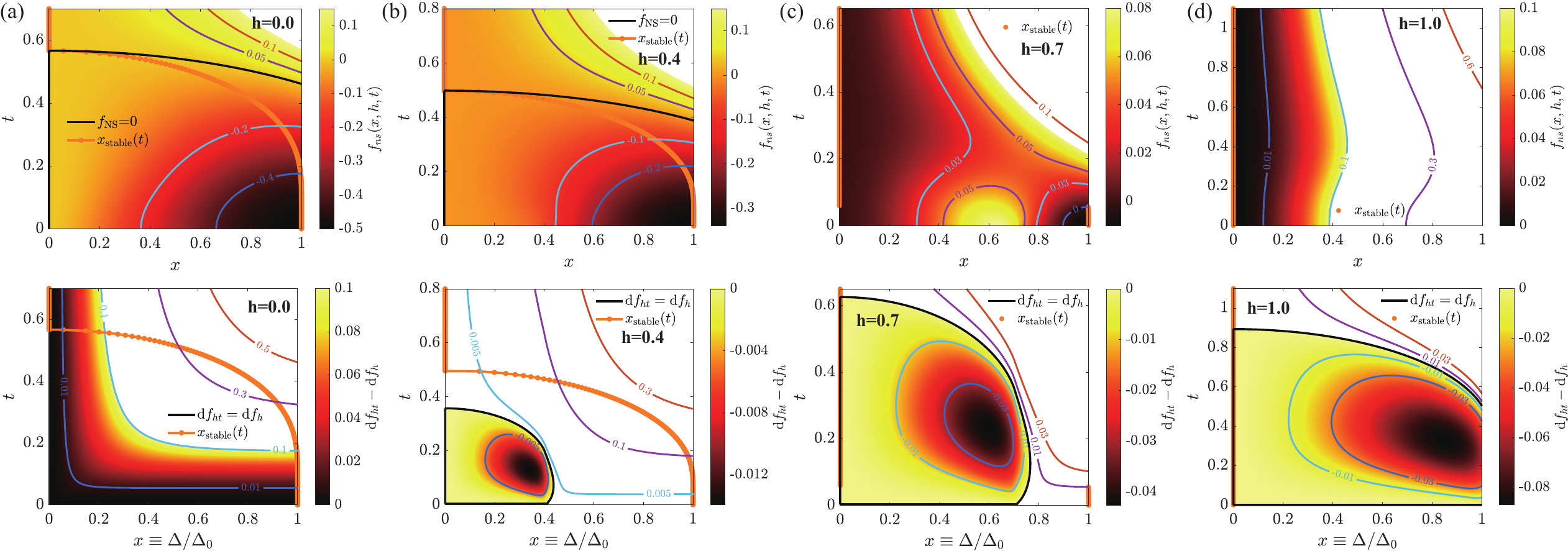}
  \caption{
  Detailed free energy analysis for one-band s-wave superconductors under different exchange fields $h$. 
  (a-d) Temperature evolution maps of $f_{ns}(x,h,t)$ and $\mathrm{d}f_{ht}-\mathrm{d}f_{h}$ for 
  (a) $h=0.0$, (b) $h=0.4$, (c) $h=0.7$, and (d) $h=1.0$.
  The case $h=0.7$ corresponds to Fig.~1(d) in the main text. 
  Color scales are slightly adjusted for clarity. Values beyond the upper color-bar limit are rendered transparent but remain traceable through contour lines. The same convention applies to subsequent figures.}
  \label{detailed_free_energy_one_band}
\end{figure}

The stable order parameter $x_{\mathrm{stable}}$ is obtained by minimizing $f_{ns}(x)$, as indicated by the orange lines or points in Fig.~\ref{detailed_free_energy_one_band}. At $h=0$, $x_{\mathrm{stable}}(t)$ reproduces the conventional BCS $\Delta(t)$ curve. The black lines denote the boundaries where the superconducting and normal states have equal free energy. These lines intersect the $x_{\mathrm{stable}}(t)$ trajectories only at the critical temperature $t_c(h)$. For small fields, $x_{\mathrm{stable}}$ decreases smoothly to zero, indicating a second-order transition (SOT). In contrast, for large $h$ (e.g., $h=0.7<h_p=0.707$), two local minima appear at $x=0$ and $x\approx1$ when $t=0$. Upon increasing temperature, the free energy near $x=1$ rises more rapidly, causing an abrupt shift of $x_{\mathrm{stable}}$ from $x=1$ to $x=0$, signaling a first-order transition (FOT), consistent with the results of Sarma~\cite{sarma1963influence}.

More insights are gained from the maps of $\mathrm{d}f_{ht}-\mathrm{d}f_{h}$. In the absence of a field ($h=0$), $\mathrm{d}f_{ht}>\mathrm{d}f_{h}$ for all $x$ and $t$, and $\mathrm{d}f_{ht}-\mathrm{d}f_{h}$ increases with temperature. Hence, purely thermal effects always suppress superconductivity. As $h$ increases, regions satisfying $\mathrm{d}f_{ht}\leq\mathrm{d}f_{h}$, indicated by solid black contours, expand progressively. The minimum of $\mathrm{d}f_{ht}-\mathrm{d}f_{h}$ occurs at $x<h$ (but close to $x\approx h$), and the condition $\mathrm{d}f_{ht}\leq\mathrm{d}f_{h}$ can extend to $x$ slightly larger than $h$. For sufficiently large fields ($h=1$), where $h>x$ for all $x$, $\mathrm{d}f_{ht}\leq\mathrm{d}f_{h}$ holds throughout the entire range, implying that thermal effects can reduce the free energy at all superconducting states. However, in one-band s-wave superconductors, temperature-induced superconductivity enhancement cannot occur because the system undergoes a FOT before entering the regime $h>x$.

Although no enhancement is observed in the one-band case, the maps of $\mathrm{d}f_{ht}-\mathrm{d}f_{h}$ clearly indicate that thermal effects suppress spin polarization more effectively than they suppress superconductivity when $h>x$, as discussed in the main text. Achieving actual enhancement therefore requires a mechanism that stabilizes superconductivity in the $h>x$ regime, such as the two-band systems examined below.

\subsection{B. Two-band free energy analysis}

The previous analysis focused on the one-band case. We now extend the discussion to two-band superconductors within the regime exhibiting temperature-induced enhancement. The parameters are chosen as $\alpha=3.0$ and $\gamma=10.0$, identical to those in Figs.~2(b) and 2(d) of the main text.

\begin{figure}[ht]
  \centering
  \includegraphics[width=0.95\linewidth]{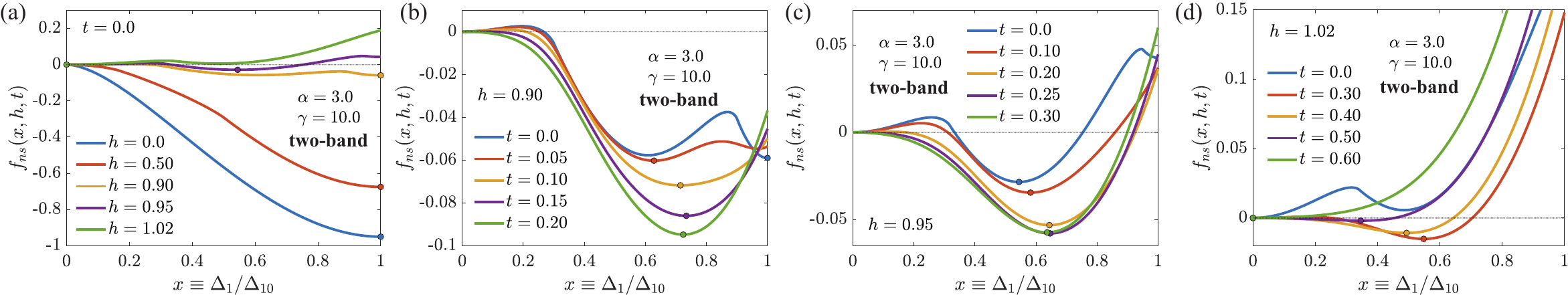}
  \caption{
  Detailed free energy landscapes for two-band s-wave superconductors with $\alpha=3.0$ and $\gamma=10.0$. 
  (a) Zero-temperature free energy landscapes $f_{ns}(x)$ for 
  $h=0.0$, $0.50$, $0.90$, $0.95$, and $1.02$.
  (b-d) Temperature-dependent landscapes of $f_{ns}(x)$ for 
  (b) $h=0.90$, (c) $h=0.95$, and (d) $h=1.02$.}
  \label{detailed_landscape_two_band}
\end{figure}

Figure~\ref{detailed_landscape_two_band}(a) shows the zero-temperature free energy landscapes for various $h$. As the exchange field increases, a partially suppressed order parameter emerges near $h\approx0.95$, and superconductivity vanishes at $h\approx1.02$. As discussed in the main text [Fig.~2(d)], high field conditions lead to non-monotonic temperature-dependent behaviors. Figures~\ref{detailed_landscape_two_band}(b-d) illustrate the detailed temperature evolution of the free energy landscapes corresponding to these cases. Notably, in certain regions of $x$, the finite-temperature free energy becomes lower than that at $t=0$, revealing a temperature-induced energy saving. To highlight this behavior, we further compute the temperature evolution maps of $f_{ns}(x,h,t)$ and $\mathrm{d}f_{ht}-\mathrm{d}f_{h}$, shown in Fig.~\ref{detailed_free_energy_two_band}.

\begin{figure}[ht]
  \centering
  \includegraphics[width=0.95\linewidth]{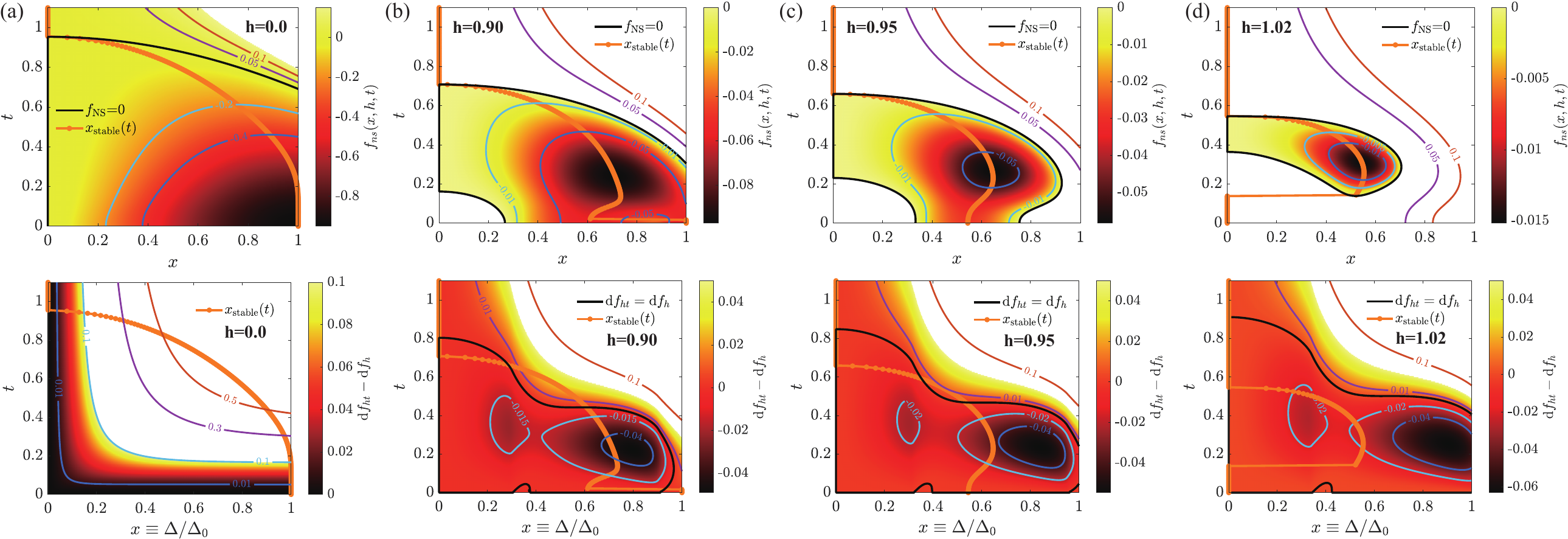}
  \caption{
  Temperature evolution maps of $f_{ns}(x,h,t)$ and $\mathrm{d}f_{ht}-\mathrm{d}f_{h}$ for two-band s-wave superconductors with $\alpha=3.0$ and $\gamma=10.0$.
  (a-d) Results for 
  (a) $h=0.0$, 
  (b) $h=0.90$, 
  (c) $h=0.95$, and 
  (d) $h=1.02$.}
  \label{detailed_free_energy_two_band}
\end{figure}

The two-band results share several qualitative similarities with the one-band case shown in Fig.~\ref{detailed_free_energy_one_band}. When $h=0$, $\mathrm{d}f_{ht}\geq\mathrm{d}f_{h}$ holds for all $x$ and $t$. As $h$ increases, the region satisfying $\mathrm{d}f_{ht}\leq\mathrm{d}f_{h}$ gradually expands. Importantly, the multiband effect allows the equilibrium order parameter to enter these regions where thermal effects reduce spin polarization more effectively than they weaken superconductivity, thereby enabling the occurrence of temperature-induced superconductivity enhancement.

\newpage
\section{IV. Temperature-induced superconductivity enhancement in highly anisotropic one-band superconductors}

The results obtained for two-band systems suggest that a partially suppressed order parameter can permit the presence of an exchange field whose energy exceeds the superconducting order parameter, thereby enabling temperature-induced superconductivity enhancement. In anisotropic one-band superconductors, the order parameter possesses angular regions with small magnitude or even nodes, which facilitate its suppression by the exchange field. Consequently, an anisotropic one-band system, such as d-wave symmetry ($\chi(\phi)=\cos2\phi$), is more promising to exhibit temperature-induced enhancement than isotropic one-band system. Furthermore, as discussed in the main text, the one-band limit can be directly derived from the two-band formulation by setting $\alpha=1$, providing a convenient benchmark to verify both the accuracy and consistency of our model. On this basis, we analyzed in detail the one-band s-wave and d-wave cases.

\begin{figure}[ht]
  \centering
  \includegraphics[width=0.7\linewidth]{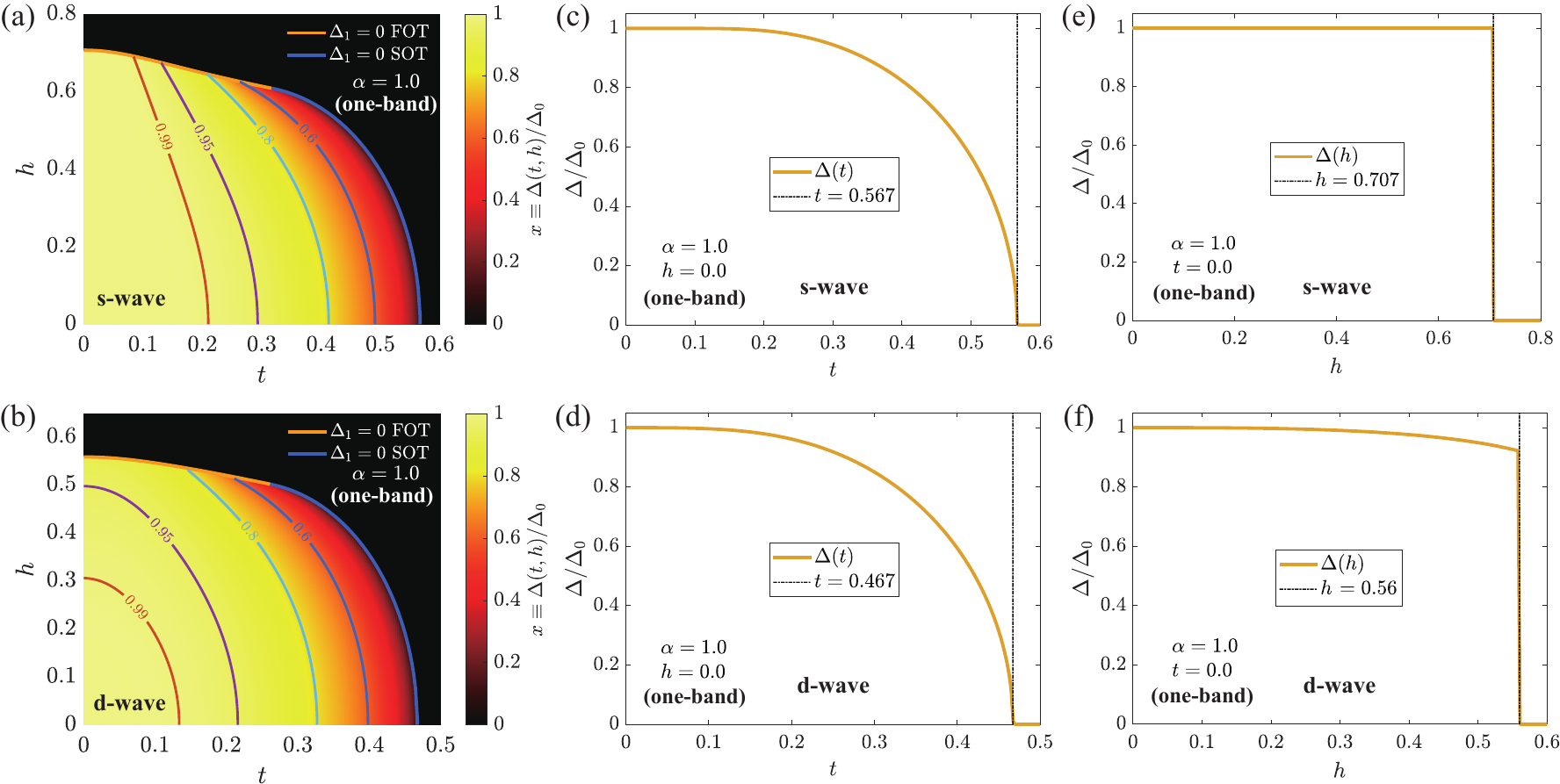}
  \caption{
  One-band s-wave and d-wave results derived from the reduced two-band model with $\alpha=1$.
  (a,b) $h$-$t$ phase diagrams for the (a) s-wave and (b) d-wave cases.
  (c,d) Temperature-dependent $x(t)$ profiles at zero field for the (c) s-wave and (d) d-wave cases.
  (e,f) Field-dependent $x(h)$ profiles at zero temperature for the (e) s-wave and (f) d-wave cases.}
  \label{h_t_one_band}
\end{figure}

The calculated s-wave $h$-$t$ phase diagram, as well as the $\Delta(t)$ and $\Delta(h)$ profiles obtained with $\alpha=1$, are in excellent agreement with previous results for isotropic one-band superconductors~\cite{sarma1963influence}. Likewise, the Pauli limit and critical temperature for the d-wave case reproduce earlier theoretical studies~\cite{sun1995impurity}, confirming the accuracy of our approach. Notably, compared with the s-wave case, the d-wave superconductor exhibits a slower decrease of $h_c$ with increasing temperature, indicating that the anisotropic pairing symmetry can more effectively promote the enhancement of $h_c$. This conclusion is further supported by the $h_{c,max}/h_{c0}$ maps for d-wave systems presented in Sec.~VI.

For superconductors with even stronger anisotropy, the system can be mapped onto an equivalent two-band system under the assumption

\begin{equation}\label{reduced_to_two_band_aniso}
\chi(\phi)=
\begin{cases}
\frac{\Delta_1}{\Delta}, & \phi \in (0,\frac{2N_1}{N_1+N_2}\pi] ,\\[4pt]
\frac{\Delta_2}{\Delta}, & \phi \in [\pi, 2\pi).
\end{cases}
\end{equation}

Moreover, any anisotropic pairing symmetry, such as d-wave like configurations (ignoring the sign change of $\chi$), can be modeled by rearranging $\chi(\phi)$ as

\begin{equation}\label{reduced_to_two_band_rearrange}
\chi_{\mathrm{d\text{-like}}}(\phi)=
\begin{cases}
\frac{\Delta_1}{\Delta}, & \phi \in (\frac{n\pi}{2},\frac{n\pi}{2}+\frac{N_1 \pi}{2(N_1+N_2)}]\equiv A,\\[4pt]
\frac{\Delta_2}{\Delta}, & \phi \notin A,
\end{cases} \quad \mathrm{where} \ n=0,1,2,3.
\end{equation}

Therefore, highly anisotropic one-band systems can potentially present the superconductivity enhancement like two-band systems.

\newpage
\section{V. Order of phase transitions in two-band superconducting systems}

As discussed in the main text [Fig.~4(b)], two-band superconductors exhibit three distinct types of behavior depending on the parameters $\alpha$ and $\gamma$. These behaviors differ in both the number and the order of their phase transitions. The first type is $\Delta_1$-dominated, occurring in the region of small $\alpha$ and relatively large $\gamma$. It exhibits a single FOT, similar to the one-band case. The second type is $\Delta_2$-dominated, found at large $\alpha$ and relatively small $\gamma$, where an insignificant SOT appears at a low field ($h \approx x$) due to the influence of $\Delta_1$, followed by a FOT at the critical field. The third type, referred to as the two-band featured regime, emerges at intermediate $\alpha$ and $\gamma$ and exhibits two FOTs: one near $h \approx x$ and another at the critical field. Notably, the SOT in $\Delta_2$-dominated systems occurs only within a narrow region of the $\alpha$-$\gamma$ parameter space. For instance, the parameters $\alpha=4.0$ and $\gamma=1.0$ used in the main text correspond to a system with two FOTs, though the transition near $h \approx x$ is weak and resembles a SOT. These parameters were chosen for clearer visualization.

To confirm the number and order of phase transitions in each case, we analyze the field dependence of the order parameter $x(h)$ and its derivative $\mathrm{d}x/\mathrm{d}h$ at zero temperature. Because of the finite numerical resolution, we use these profiles as diagnostics: a FOT is identified by a discontinuity in $x(h)$ accompanied by a divergent peak in $\mathrm{d}x/\mathrm{d}h$, while a SOT manifests as a continuous $x(h)$ with an abrupt change in $\mathrm{d}x/\mathrm{d}h$.

\begin{figure}[ht]
  \centering
  \includegraphics[width=0.95\linewidth]{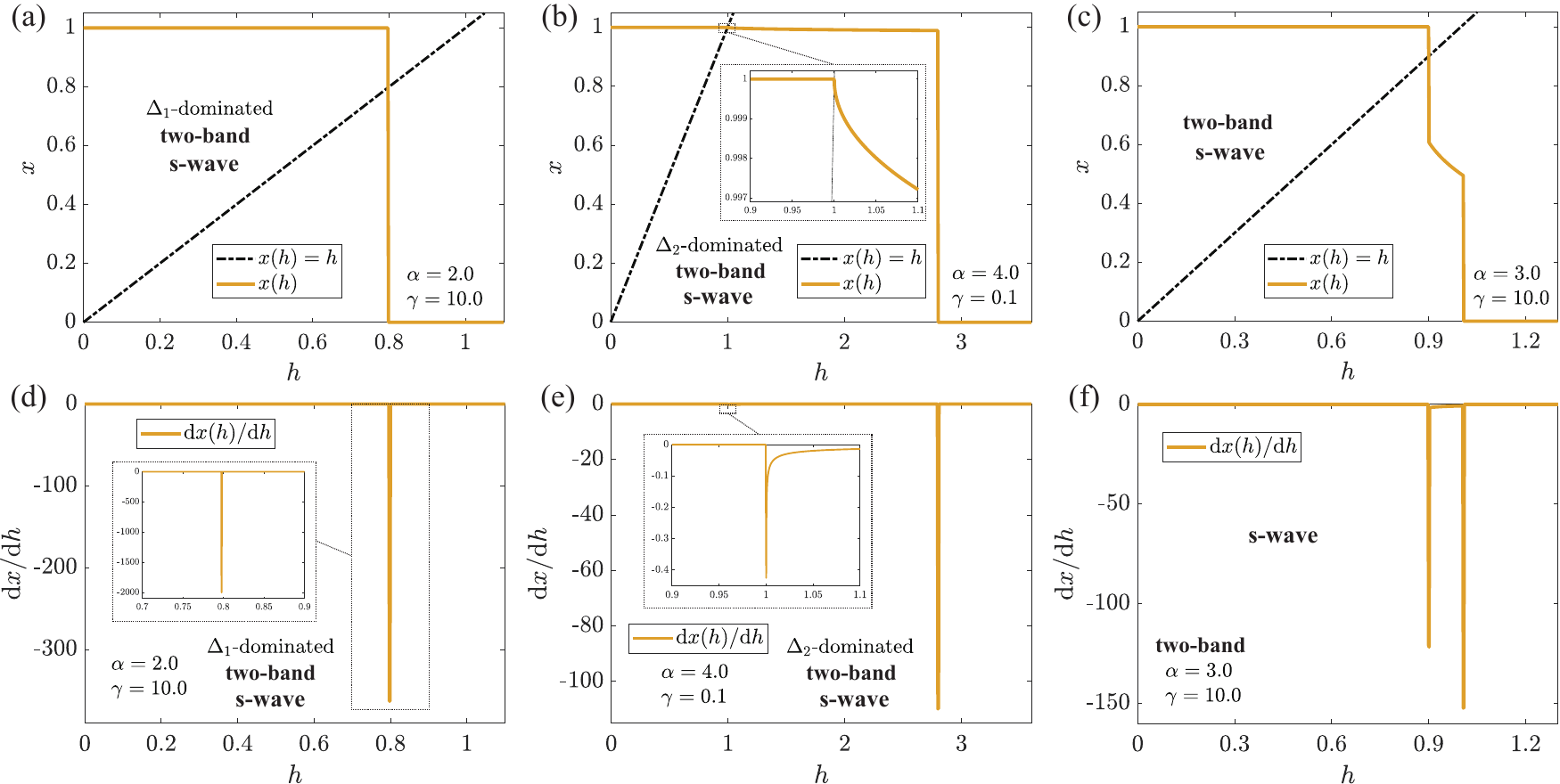}
  \caption{
  Order and number of phase transitions in representative two-band superconductors at zero temperature. 
  (a-c) Zero-temperature order-parameter profiles $x(h)$ for 
  (a) $\Delta_1$-dominated, (b) $\Delta_2$-dominated, and (c) two-band featured systems.
  (d-f) Corresponding derivatives $\mathrm{d}x/\mathrm{d}h$ for (a-c). 
  Insets in panels (b) and (d,e) show enlarged views near the transition points with smaller field steps, highlighting the features of phase transitions. 
  The parameters are identical to those in Fig.~4(b) of the main text, except for (b) and (e).}
  \label{order_identified_phase_transitions}
\end{figure}

As shown in Fig.~\ref{order_identified_phase_transitions}, $\Delta_1$-dominated systems exhibit a discontinuity in $x(h)$ and a pronounced peak in $\mathrm{d}x/\mathrm{d}h$ only at the critical field, confirming the presence of a FOT. Although $\mathrm{d}\Delta_1/\mathrm{d}h$ does not diverge numerically due to finite step size, the peak amplitude increases proportionally with smaller step size, consistent with the feature of FOT. In contrast, $\Delta_2$-dominated systems display a smooth $x(h)$ but an abrupt change in $\mathrm{d}x/\mathrm{d}h$ near $h \approx x$, confirming the existence of an SOT. Two-band featured systems show two distinct discontinuities in $x(h)$ and two corresponding peaks in $\mathrm{d}x/\mathrm{d}h$, indicating two FOTs.

In principle, a full phase diagram distinguishing these three behaviors can be mapped in the $\alpha$-$\gamma$ parameter space by identifying characteristic boundaries such as $x(h_{c0}^{-})=1$ in Fig.~4(a), which separates $\Delta_1$-dominated regions from the others. Furthermore, the temperature dependence of transition order can also be examined: thermal effects may convert both FOT and SOT into continuous crossovers. A detailed analysis of these finite-temperature effects will be presented in a forthcoming publication.

\newpage
\section{VI. Superconductivity enhancement in two-band d-wave superconductors}

The phenomenon of temperature-induced superconductivity enhancement in two-band systems can be extended to many other pairing symmetries beyond s-wave, such as d-wave. Here, we present representative results demonstrating $h_c$ enhancement in the two-band d-wave case and calculate the corresponding $h_{c,max}/h_{c0}$ map in the $\alpha$-$\gamma$ parameter space. Because of the high computational cost, the map was evaluated on a coarse $20\times20$ grid. For direct comparison, the d-wave results are contrasted with the s-wave counterparts obtained under the same parameters.

\begin{figure}[ht]
  \centering
  \includegraphics[width=0.8\linewidth]{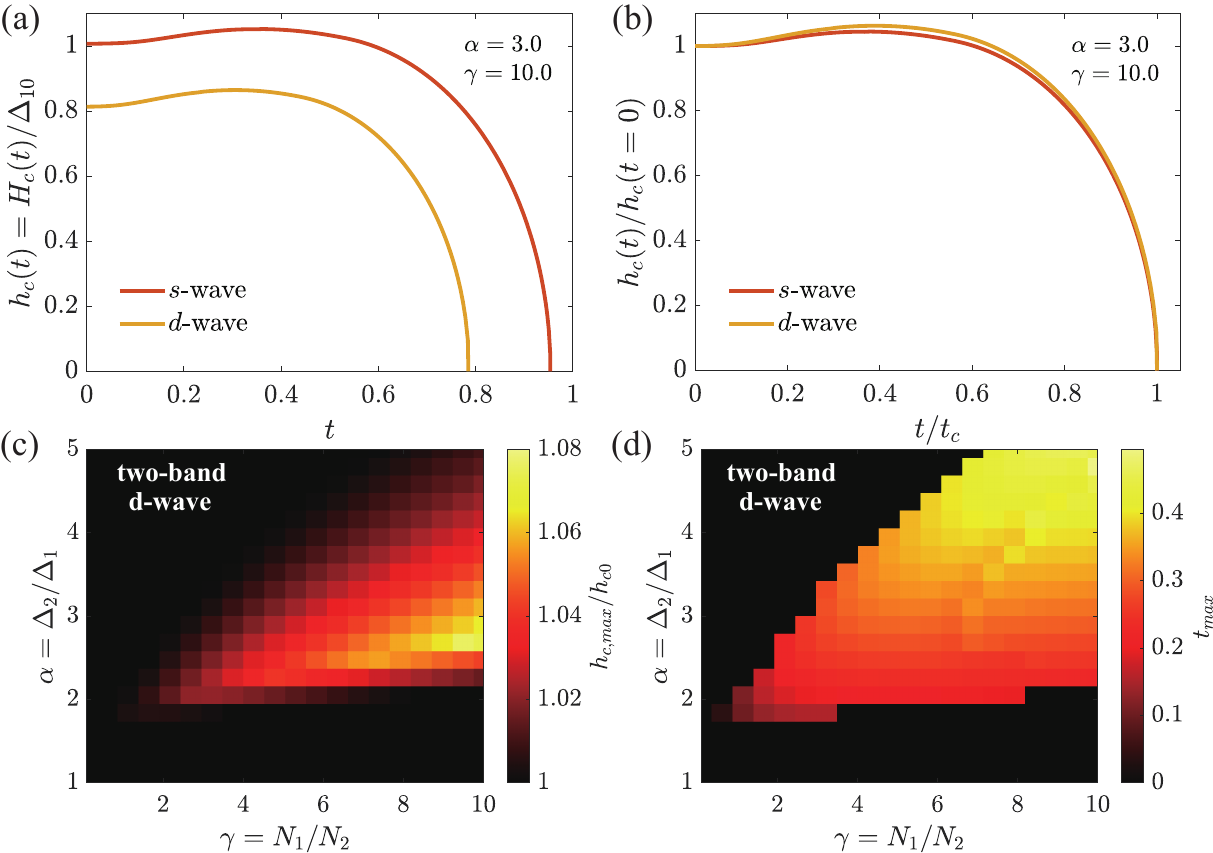}
  \caption{
  Superconductivity enhancement in two-band d-wave systems.
  (a,b) Comparison of 
  (a) $h_c(t)$ and 
  (b) normalized $h_c(t)/h_c(0)$ curves for s-wave and d-wave cases with identical parameters ($\alpha=3$, $\gamma=10.0$).
  (c,d) Maps of 
  (c) $h_{c,max}/h_{c0}$ and 
  (d) $t_{max}$ for the d-wave case in the $\alpha$-$\gamma$ parameter space.
  Color scales in (c) and (d) denote the amplitudes of $h_{c,max}/h_{c0}$ and $t_{max}$, respectively.}
  \label{d_wave_enhancement}
\end{figure}

As shown in Fig.~\ref{d_wave_enhancement}(a), the s-wave and d-wave cases exhibit similar overall trends in $h_c(t)$. However, small but distinct differences appear in the normalized curves $h_c(t)/h_c(0)$ [Fig.~\ref{d_wave_enhancement}(b)]. Compared with the s-wave case, the d-wave superconductor displays a more pronounced enhancement of $h_c$ at finite temperature, which is further confirmed by the $h_{c,max}/h_{c0}$ map.

The $h_{c,max}/h_{c0}$ distribution for the d-wave case closely resembles that of the s-wave system, with enhancement ($h_{c,max}/h_{c0}>1$) occurring primarily in the middle-right region of the $\alpha$-$\gamma$ space, where both parameters are comparable and two-band effects are most evident. Notably, the d-wave system exhibits both a larger area of $h_c$ enhancement and higher maximum values of $h_{c,max}/h_{c0}$, consistent with the discussion in Sec.~IV. This stronger enhancement arises because the anisotropic d-wave gap contains more regions with small order-parameter amplitude, allowing a stronger partial suppression of the order parameter $x$ by the exchange field and enabling relatively higher $h_c$. In addition, the temperature corresponding to the maximal critical field, $t_{max}$, also shows a slight upward shift compared with the s-wave case.

\newpage
\section{VII. Additional temperature-related non-monotonic behaviors}

As discussed in the main text, various temperature-induced non-monotonic superconducting behaviors can arise in two-band systems. Here, we present two additional examples beyond those shown in the main text to further demonstrate the variety of these behaviors. The examples are deliberately taken for both s-wave and d-wave cases to illustrate that such behaviors are not limited to a specific pairing symmetry.

\subsection{A. Non-monotonic behavior in two-band s-wave systems}

We first consider the two-band s-wave case. As indicated in Fig.~3(c) of the main text, the $h_c$ enhancement phenomenon can extend slightly into the $\Delta_1$-dominated region, where a distinct temperature-related non-monotonic behavior emerges.

\begin{figure}[ht]
  \centering
  \includegraphics[width=0.65\linewidth]{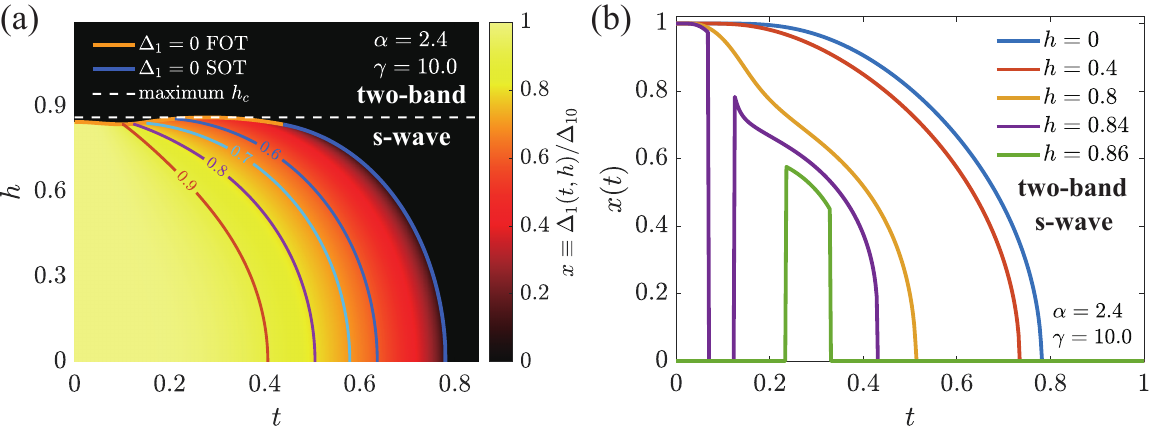}
  \caption{
  Another temperature-related non-monotonic behavior in a two-band s-wave system with $\alpha=2.4$ and $\gamma=10.0$. 
  (a) $h$-$t$ phase diagram. 
  (b) Temperature dependence of the order parameter $x(t)$ at several exchange fields. 
  The color scale in (a) indicates the magnitude of the order parameter $x$, and contour lines highlight the non-monotonic features.}
  \label{s_wave_non_monotonic}
\end{figure}

As shown in Fig.~\ref{s_wave_non_monotonic}(a), the $h_c(t)$ curve exhibits a complex non-monotonic evolution: $h_c$ first decreases, then increases, and finally decreases again with rising temperature. This sequence produces a double-extrema structure in $h_c(t)$, which is further illustrated by the $x(t)$ profiles at $h=0.84$ in Fig.~\ref{s_wave_non_monotonic}(b).

\subsection{B. Non-monotonic behavior in two-band d-wave systems}

We next turn to the two-band d-wave case. Near the center of the $h_c$ enhanced region in parameter space, another form of non-monotonic behavior appears.

\begin{figure}[ht]
  \centering
  \includegraphics[width=0.65\linewidth]{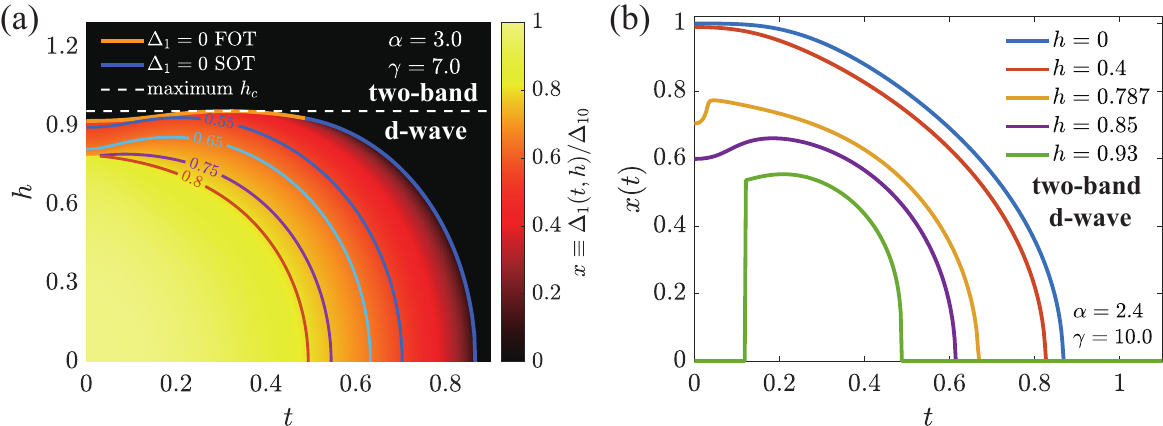}
  \caption{
  Another temperature-related non-monotonic behavior in a two-band d-wave system with $\alpha=3.0$ and $\gamma=7.0$. 
  (a) $h$-$t$ phase diagram. 
  (b) Temperature dependence of the order parameter $x(t)$ at several exchange fields. 
  The color scale in (a) represents the order-parameter amplitude $x$, and contour lines emphasize non-monotonic superconducting features.}
  \label{d_wave_non_monotonic}
\end{figure}

In Fig.~\ref{d_wave_non_monotonic}(a), the field corresponding to the first FOT exhibits a subtle upward shift with increasing temperature, giving rise to a distinct non-monotonic behavior in the low-field region near $h\approx x$. This effect is visualized more clearly in the $x(t)$ profile at $h=0.787$ [Fig.~\ref{d_wave_non_monotonic}(b)].

These temperature-related non-monotonic phenomena are not restricted to s-wave or d-wave pairing. In general, two-band superconductors with arbitrary pairing symmetry can exhibit similar behaviors when appropriate combinations of $\alpha$ and $\gamma$ parameters are selected.

\end{widetext}

\end{document}